\documentclass[article]{IEEEtran}
\usepackage{tabularx}
\usepackage[T1]{fontenc}
\usepackage{graphicx}
\usepackage{algorithm}
\usepackage{algpseudocode}
\usepackage{amsmath,amssymb,amsfonts}
\usepackage{array}
\usepackage{textcomp}
\usepackage[font=small]{caption}
\usepackage{paralist}
\usepackage{lettrine}
\usepackage{cite}
\usepackage{xcolor}
\usepackage{cancel}
\usepackage{float}
\usepackage{booktabs}
\usepackage{subfig}
\usepackage{stfloats}
\usepackage{url}
\usepackage{lipsum}
\usepackage{enumitem}
\usepackage{accents}
\usepackage{mathtools}
\usepackage{cuted}
\usepackage{makecell}
\usepackage{ifthen}
\usepackage{comment}
\usepackage{acronym}
\usepackage{multicol}
\usepackage{multirow}
\usepackage{authblk}
\usepackage{fixltx2e}
\usepackage{glossaries-extra}
\setabbreviationstyle[acronym]{long-short}
\newacronym{AoI}{AoI}{angle of incidence}
\newacronym{AoR}{AoR}{angle of reflection}
\newacronym{AoA}{AoA}{angle of arrival}
\newacronym{SDR}{SDR}{semidefinite relaxation}
\newacronym{AoD}{AoD}{angle of departure}
\newacronym{EM}{EM}{electromagnetic}
\newacronym{SINR}{SINR}{signal-to-interference-and-noise ratio}
\newacronym{RIS}{RIS}{reconfigurable intelligent surface}
\newacronym{IRS}{IRS}{intelligent reflective surface}
\newacronym{GA}{GA}{genetic algorithm}
\newacronym{CIRS}{CIRS}{curved intelligent reflective surface}

\newacronym{SNR}{SNR}{signal to noise ratio}

\newacronym{V2V}{V2V}{vehicle-to-vehicle}
\newacronym{SDP}{SDP}{semidefinite programming}
\newacronym{PDF}{PDF}{probability density function}
\newacronym{CDF}{CDF}{cumulative 
distribution function}
\newacronym{TMA}{TMA}{tassled modules architecture} 

\newacronym{CEMS}{CEMS}{curved EMS} 

\newacronym{SP-CEMS}{SP-CEMS}{static passive curved EMS} 

\newacronym{EMS}{EMS}{Electromagnetic skins}

\newacronym{AO}{AO}{alternating optimization }
\newacronym{SE}{SE}{spectral efficiency }

\newacronym{ECDF}{ECDF}{empirical cumulative density function}

\newacronym{SmSk}{SmSk}{smart skins}
\newacronym{FF}{FF}{far-field}

\newcommand{\widebar}[1]{\mkern1.5mu\overline{\mkern-1.5mu#1\mkern-1.5mu}\mkern1.5mu}
\newcommand{\RN}[1]{%
  \textup{\uppercase\expandafter{\romannumeral#1}}%
}

\def\BibTeX{{\rm B\kern-.05em{\sc i\kern-.025em b}\kern-.08em
    T\kern-.1667em\lower.7ex\hbox{E}\kern-.125emX}}

\title{Optimizing Curved EM Skins for Opportunistic Relaying in Vehicular Networks

\author{Reza Aghazadeh Ayoubi,~\IEEEmembership{Graduate Student Member,~IEEE},
 Silvia~Mura,~\IEEEmembership{Member,~IEEE},
 Dario~Tagliaferri,~\IEEEmembership{Member,~IEEE},
 Marouan~Mizmizi,~\IEEEmembership{Member,~IEEE}, Umberto~Spagnolini,~\IEEEmembership{Senior Member,~IEEE}
\thanks{The authors are with the  Department of Electronics, Information and Bioengineering (DEIB) of Politecnico di Milano, 20133 Milan, Italy  (e-mail: [reza.aghazadeh, silvia.mura, dario.tagliaferri, marouan.mizmizi, umberto.spagnolini]@polimi.it }
}
\thanks{This work was supported by the European Union under the Italian National Recovery and Resilience Plan (NRRP) of NextGenerationEU, partnership on “Telecommunications of the Future” (PE00000001 - program “RESTART”).}
}

\begin{document}
\maketitle

\begin{abstract}
Electromagnetic skins (EMSs) are recognized for enhancing communication performance, spanning from coverage to capacity. While much of the scientific literature focuses on reconfigurable intelligent surfaces that dynamically adjust phase configurations over time, this study takes a different approach by considering low-cost static passive \gls{CEMS}s. These are pre-configured during manufacturing to conform to the shape of irregular surfaces, e.g., car doors, effectively transforming them into anomalous mirrors. This design allows vehicles to serve as opportunistic passive relays, mitigating blockage issues in vehicular networks.
This paper delves into a novel design method for the phase profile of \gls{CEMS} based on coarse a-priori distributions of incident and reflection angles onto the surface, influenced by vehicular traffic patterns. A penalty-based method is employed to optimize both the average \gls{SE} and average coverage probability, and it is compared against a lower-complexity and physically intuitive modular architecture, utilizing a codebook-based discrete optimization technique. Numerical results demonstrate that properly designed \gls{CEMS} lead to a remarkable improvements in average \gls{SE} and coverage probability, namely when the direct path is blocked.
\end{abstract}

\section{Introduction}
Metasurfaces have become a focal point of research in the field of sixth-generation wireless systems (6G), especially concerning high-frequency wireless communications like millimeter-wave (mmWave) ($30-100$ GHz) and sub-THz frequencies ($>100$ GHz). These surfaces offer vast potential across a spectrum of applications, encompassing communication and sensing \cite{SurveyDiRenzo}. Functioning as two-dimensional structures, metasurfaces possess remarkable abilities to manipulate the propagation characteristics of electromagnetic (EM) waves. This manipulation is achieved through precise control of the amplitude and/or phase of the reflection coefficient of the surface, utilizing principles akin to the generalized Snell's law \cite{GSnell}. In practical applications, discretized versions of metasurfaces, namely \textit{EM skins}, are synthesized, adhering to the generalized sheet transition condition \cite{GSTC}. EM skins are made of sub-wavelength-sized elements (meta-atoms), whose phases and possibly amplitudes of the reflection/transmission coefficient can be tuned according to needs, e.g., for anomalous reflection/refraction, focusing, etc. \cite{SurveyDiRenzo}. 
\gls{RIS} are the prevalent and extensively studied type of EM skins in contemporary literature \cite{9475160}. \gls{RIS} enables dynamic reconfiguration of the phase of each meta-atom to alter the device's behavior over time. This capability finds applications such as dynamic beam alignment in communication \cite{haghshenas2023parametric} or scanning for sensing purposes \cite{bellini2024multiview}. When the amplitude of \gls{RIS} elements can also be adjusted, often achieved by applying an amplification gain to the incident signal, they are termed as active \gls{RIS}. The latter have been investigated for both localization \cite{ARIS_Localization} and radar applications \cite{ARIS_Radar}.

\subsection*{Related works}

High-frequency communications encounter significant challenges, including severe path and penetration losses, leading to diminished coverage and blockage issues\cite{linsalata2022map}. These factors pose notable constraints on widespread implementations.
Therefore, initial research on \gls{RIS} for communication has focused on addressing signal-to-noise ratio (SNR) concerns \cite{OPT_SNR_1,OPT_SNR_4}, optimizing sum-rate performance \cite{OPT_RATE_2,OPT_RATE_3}, and enhancing coverage in millimeter-wave (mmWave) networks \cite{COV_Analysis_1,COV_Analysis_3}.
The primary objective of the aforementioned studies is to assess the coverage probability with and without RISs, demonstrating the advantages of employing a pre-defined (and optimal) phase shift design. Other research endeavors have targeted phase shift design specifically for coverage improvement \cite{COV_OPT_1,COV_OPT_2}.
In \cite{COV_OPT_1} and \cite{COV_OPT_2}, the orientation and location of the \gls{RIS}s are optimized to maximize the coverage probability, while the phase shifts of the RIS elements are assumed to be dynamically optimized. Incorporating statistical user position information has been explored in a few studies, such as \cite{OPT_RATE_2,OPT_RATE_12}, primarily for sum-rate optimization.

Within the vehicle-to-vehicle (V2V) communication domain, some studies focused on the physical design of the RIS, see e.g., \cite{V2V_element_1}. From the system level perspective, various aspects of RIS-assisted communication, including SNR enhancement, phase shift design, and user scheduling, have been examined across multiple works as \cite{V2V_1,V2V_2}. These latter studies typically consider the RIS as fixed and located at a roadside unit or onto buildings, serving vehicles.

%

Despite being recognized as a promising technology, RISs face significant limitations in practical implementation. These challenges encompass relatively high manufacturing costs \cite{Danilo_SmartEnv}, power consumption concerns, and the need to dynamically reconfigure the phase gradient of the surface due to the channel's variability. This requirement for periodic channel estimation incurs known costs in terms of signaling overhead and spectral efficiency \cite{haghshenas2023parametric}. 
Furthermore, the planar nature of RIS makes installation on curved surfaces. challenging. A pioneering approach aimed at addressing such challenges involves static passive pre-configured EM skins \cite{smsk_electronics,SmartSkin}. These surfaces are pre-configured during the manufacturing process, thus enabling cost-effective production. Various techniques, such as altering surface thickness utilizing polymer materials \cite{Polymer}, or printing on adhesive stickers \cite{InkJetPrinted,Transparent}, can be employed for their synthesis. However, pre-configured EM skins cannot dynamically align to the specific positions or angles of interest, which may cause a loss in received signal power. Nevertheless, suitable strategies are available for optimally pre-configuring such surfaces. By doing so, the loss of received signal power can be minimized \cite{SmartSkin,SMSK_Footprint}.
%
In our prior work \cite{CIRS_TWC}, we introduced the concept of static passive curved EM skins (CEMS). Unlike RISs, which are both reconfigurable and flat, the \gls{CEMS} are non-reconfigurable and can be lodged to objects with arbitrarily shaped facades. This feature allows for the installation of the EM skin on the lateral sides of vehicles, facilitating direct V2V communication. It was demonstrated in \cite{CIRS_TWC} that through the design of the passive phase gradient of such surfaces, it is possible to achieve the specular reflection behavior of a flat surface. The curved bodies of cars equipped with \gls{CEMS} can enhance the received SNR when the direct path between a transmitter and receiver is obstructed.

\begin{figure*}[t!]
    \centering\subfloat{\includegraphics[width=0.95\textwidth]{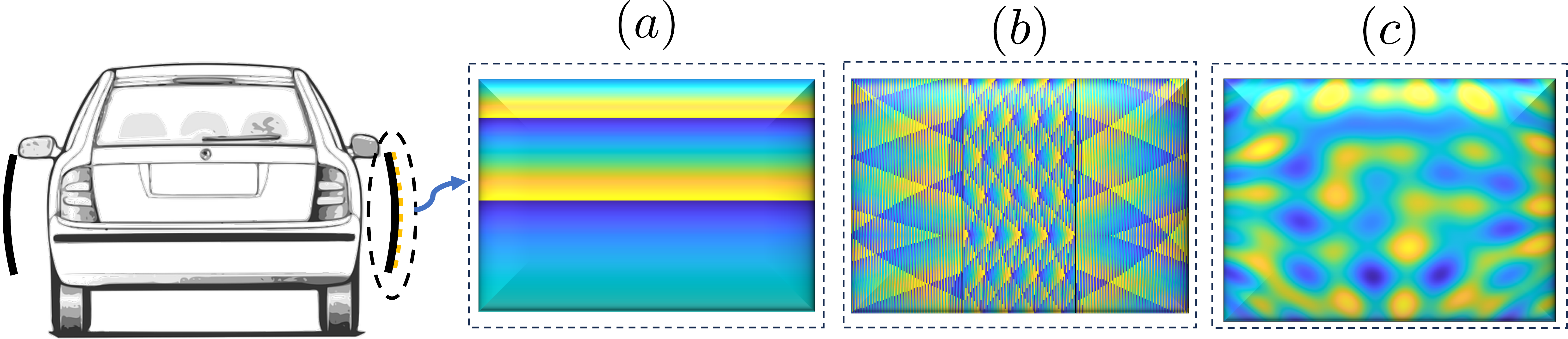}}
    \caption{Conceptual illustration of CEMS phase gradient, designed (a) for specular reflection \cite{CIRS_TWC}, (b) for anomalous reflection, employing the structured optimization method in Section \ref{sec:modular_design} (3 modules), and (c) for anomalous reflection, employing the unstructured optimization method in Section \ref{sec:Opt_Unstructured}.}
    \label{fig:conceptual_illustration}
\end{figure*}

\subsection*{Contributions}

This paper focuses on the design of CEMS for V2V. We design the phase configuration of the CEMS to maximize either \gls{SE} or coverage probability, specifically exploiting the statistical distribution of the position of the vehicles, which is linked to the probability density function (pdf) of angles of incidence/reflection (AoI and AoR) in typical traffic scenarios. Remarkably, no statistical position information for the users is used in the available literature on static passive EM skins. Existing works target planar surfaces installed on buildings for wireless communication as in \cite{SMSK_Footprint,SMSK_genetic}, thus their aim is to design a specific reflection pattern by solving an inverse problem, rather than directly optimizing a communication metric. Other works, such as~\cite{OPT_SNR_1}, rely on \textit{unstructured design methods}, such as \gls{SDR}, to optimize the phase of each element of the surface independently, without adhering to a closed-form rule for the pattern. This latter approach guarantees the highest flexibility is choosing the phase pattern at the price of being poorly scalable in terms of computational complexity, limiting the application to small surfaces.



%

The contributions of the paper are as follows:

\begin{itemize}
    \item We introduce an innovative unstructured design method, that accounts for the statistical distributions of vehicles and the associated pdf of AoI and AoR. Through numerical demonstrations, we illustrate that the outcomes in terms of \gls{SE} and coverage probability outperform those achieved by the traditional \gls{SDR} method in \cite{OPT_SNR_1}.
    \item To face the inherent complexity of the aforementioned unstructured method, we propose a modular architecture along with a \textit{structured} design, utilizing a highly efficient discrete angular codebook based on a pre-defined model for the phase gradient. Through numerical simulations, we show that this design method, can achieve performance comparable to that of the unstructured penalty-based method in terms of \gls{SE} and coverage probability, all while maintaining lower complexity. This capability enables the design of very large-sized \gls{CEMS}s.
    \item We conduct a system-level V2V communication simulation within a multi-lane highway scenario, considering blockage probability. Through this simulation, we illustrate that the optimized CEMS can substantially enhance both the average \gls{SE} and coverage probability.
\end{itemize}

It is worth noting that the simulations in this paper are specifically tailored for V2V scenarios, and a conceptual illustration showcasing the phase gradient of \gls{CEMS} with various configurations, illustrated later, is depicted in Fig. \ref{fig:conceptual_illustration}. Nevertheless, the methods introduced hold broader validity and can be applied to synthesize fully passive metasurfaces for arbitrary networks or indoor applications, provided that the PDFs of the angles are known.

\subsection*{Organization}
In Sec. \ref{sect:system_model} system and channel model is re-presented from \cite{CIRS_TWC}, to make the paper self-consistent. In Sec. \ref{sect:SPCEMS_config}, the general concept of phase configuration for \gls{CEMS} is discussed. In Sec.\ref{sec:Opt_Unstructured} the unstructured optimization method for the \gls{SE} and coverage is discussed. In section \ref{sec:modular_design}, the structured optimization method for \gls{SE} and coverage is presented.
Section \ref{sect:numerical_results} is the simulations and numerical results, and finally the paper is concluded in Section \ref{sect:conclusion}.
\subsection*{Notation}
Bold upper- and lower-case letters describe matrices and column vectors. The $(i,j)$-th entry of matrix $\mathbf{A}$ is denoted by $[\mathbf{A}]_{ij}$. Matrix transposition, conjugation, conjugate transposition, Frobenius norm and nuclear norm are indicated respectively as $\mathbf{A}^T$, $\mathbf{A}^{*}$, $\mathbf{A}^H$, $\|\mathbf{A}\|_F$ and $\|\mathbf{A}\|_*$. $\mathrm{tr}\left(\mathbf{A}\right)$ extracts the trace of $\mathbf{A}$. $\mathrm{diag}(\mathbf{A})$ denotes the extraction of the diagonal of $\mathbf{A}$, while $\mathrm{diag}(\mathbf{a})$ is the diagonal matrix given by vector $\mathbf{a}$. $\mathbf{I}_n$ is the identity matrix of size $n$. With  $\mathbf{a}\sim\mathcal{CN}(\boldsymbol{\mu},\mathbf{C})$ we denote a multi-variate circularly complex Gaussian random variable $\mathbf{a}$ with mean $\boldsymbol{\mu}$ and covariance $\mathbf{C}$. $\mathbb{E}[\cdot]$ is the expectation operator, while $\mathbb{R}$ and $\mathbb{C}$ stand for the set of real and complex numbers, respectively.

\section{Motivation and Practical Example}

\begin{figure}[t!]
    \centering
    \subfloat[CEMS with specular reflection as in \cite{CIRS_TWC}]{\includegraphics[width=0.49\textwidth]{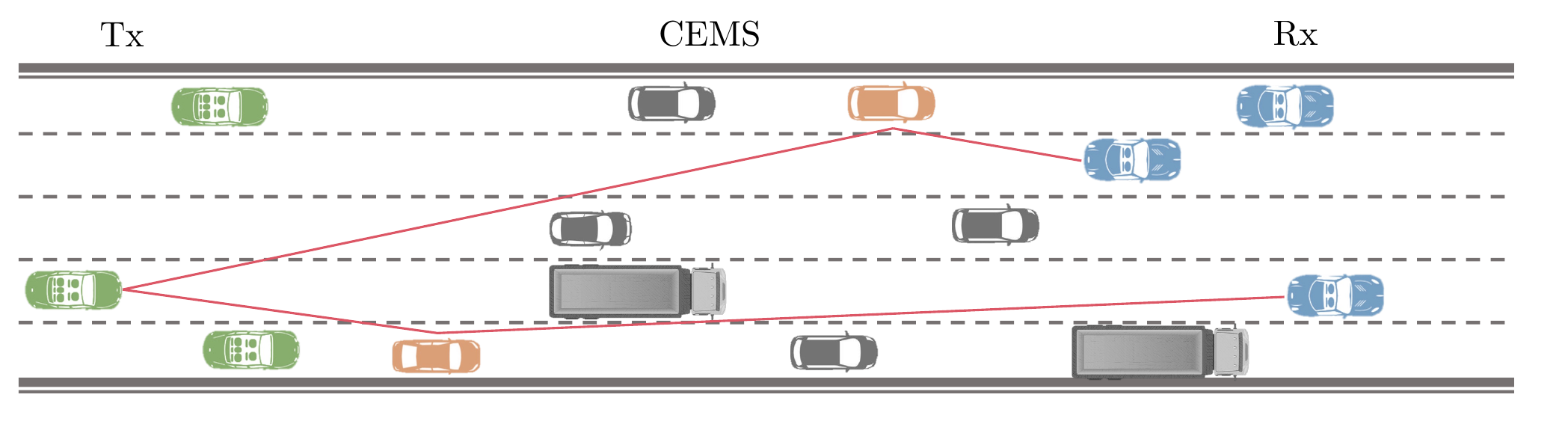}\label{subfig:scenario_specular}}\\
    \subfloat[Proposed CEMS with optimized reflection]{\includegraphics[width=0.49\textwidth] {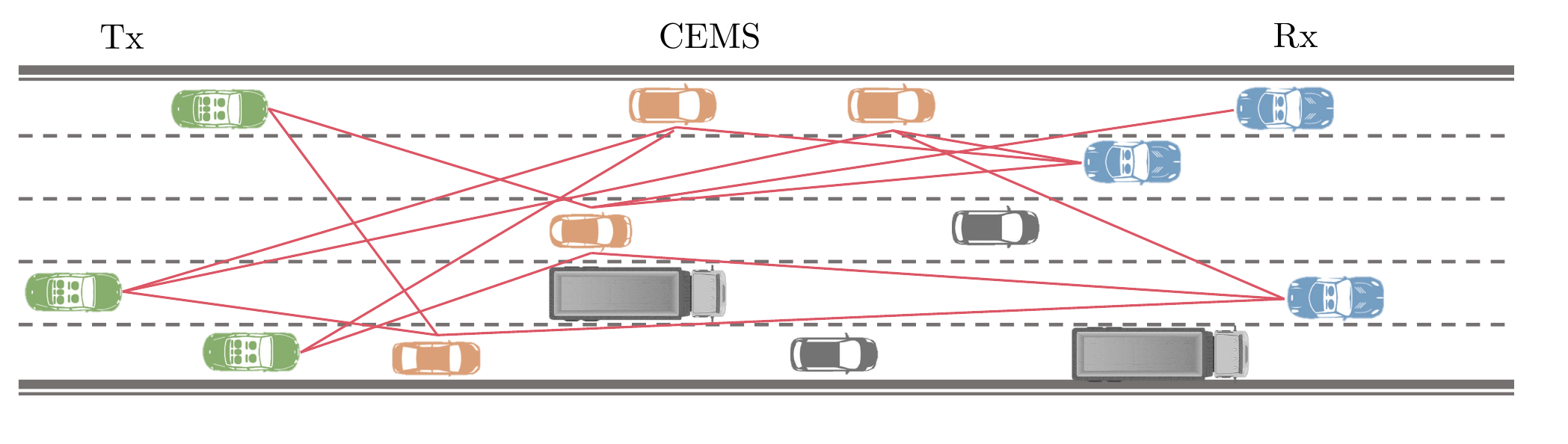}\label{subfig:scenario_Optimized}}
    \caption{V2V communication in a multi-lane highway scenario;  possible relay(s), equipped with: (a) specular reflecting CEMS, and (b) optimized reflection CEMS.}\label{fig:MultiLaneScenario}
\end{figure}

The CEMS, as proposed in \cite{CIRS_TWC}, compensates for the geometric shape of car sides and allows for a specular reflection. As a result, only a small number of vehicles can benefit from CEMS as relays, as shown in Fig.\ref{subfig:scenario_specular}. However, the novel design of this paper, offers a wider range of vehicles that can benefit from the CEMS relaying capabilities. This is due to an optimized design dictated by the distribution of vehicles in the highway scenario, as seen in Fig.\ref{subfig:scenario_Optimized}. 

To provide better motivation for the proposed work, let us consider the example shown in Fig.\ref{fig:MultiLaneScenario}. This represents a snapshot of traffic on a highway in which the candidate transmitting (Tx) vehicles are marked in green, the potential receiving (Rx) vehicles are marked in blue, and the vehicles equipped with CEMS and capable of acting as relays are marked in orange. When CEMS are designed to compensate for the geometrical shape of the surface on which they are lodged enabling specular reflections \cite{CIRS_TWC}, only certain vehicles can be in favorable positions to act as relays. 
In the given example depicted in Fig.\ref{subfig:scenario_specular}, only two candidates are available, resulting in a connectivity\footnote{The \textit{algebraic connectivity} is the second smallest eigenvalue of the Laplacian matrix of a graph, and it measures how much a network is connected. See \cite{linsalata2022map} for more details.} of $4\%$. On the other hand, the proposed solution, through an optimized design, can compensate for the geometry of the installation surface and cover more angles. This can be seen in the second figure, where four vehicles are eligible to act as relay candidates, leading to a connectivity of $33\%$ in the given example depicted in Fig. \ref{subfig:scenario_Optimized}.

\begin{figure}[!t]
    \centering
    \includegraphics[width=0.45\textwidth]{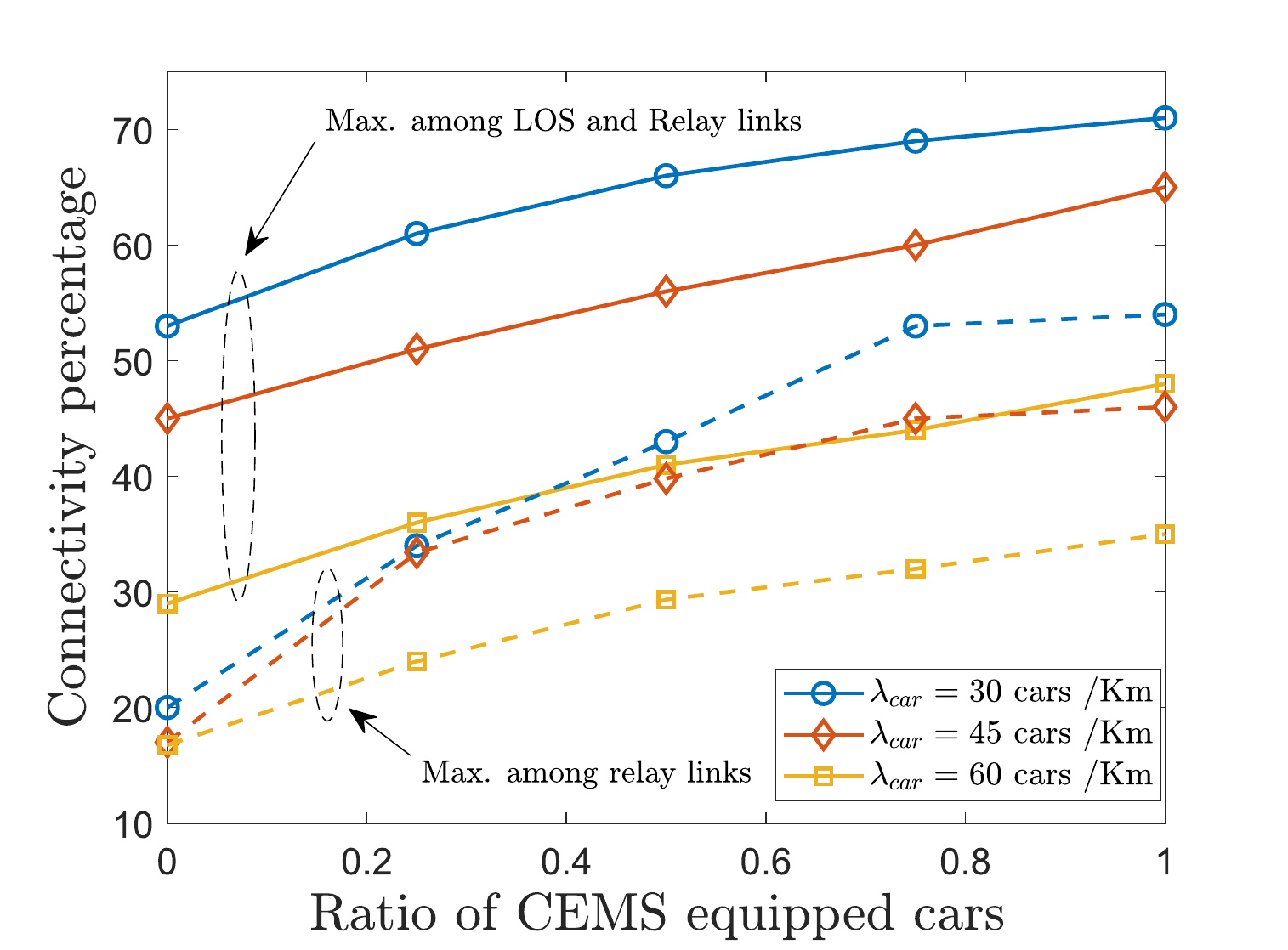}
    \caption{Network connectivity in V2V highway scenario varying the ratio of CEMS-equipped cars, i.e., the CEMS technology penetration level.}\label{fig:connectivity}
\end{figure}

The effectiveness of the proposed solution relies on the number of vehicles equipped with CEMS technology, i.e., technology penetration level. Fig. \ref{fig:connectivity} illustrates the impact of penetration level on network connectivity. The solid lines on the graph show network connectivity when the system selects the strongest signal between the direct link (considering possible blockage loss) and the relayed links. The dashed lines, instead, display network connectivity when the system considers only the relayed links. As the graph indicates, CEMS technology significantly impacts network connectivity in low-traffic density conditions. In such a scenario, having all vehicles equipped with CEMS technology can enhance network connectivity by more than $20\%$.


\section{System Model}\label{sect:system_model}

\begin{figure*}[t!]
    \centering
    \includegraphics[width=0.75\textwidth]{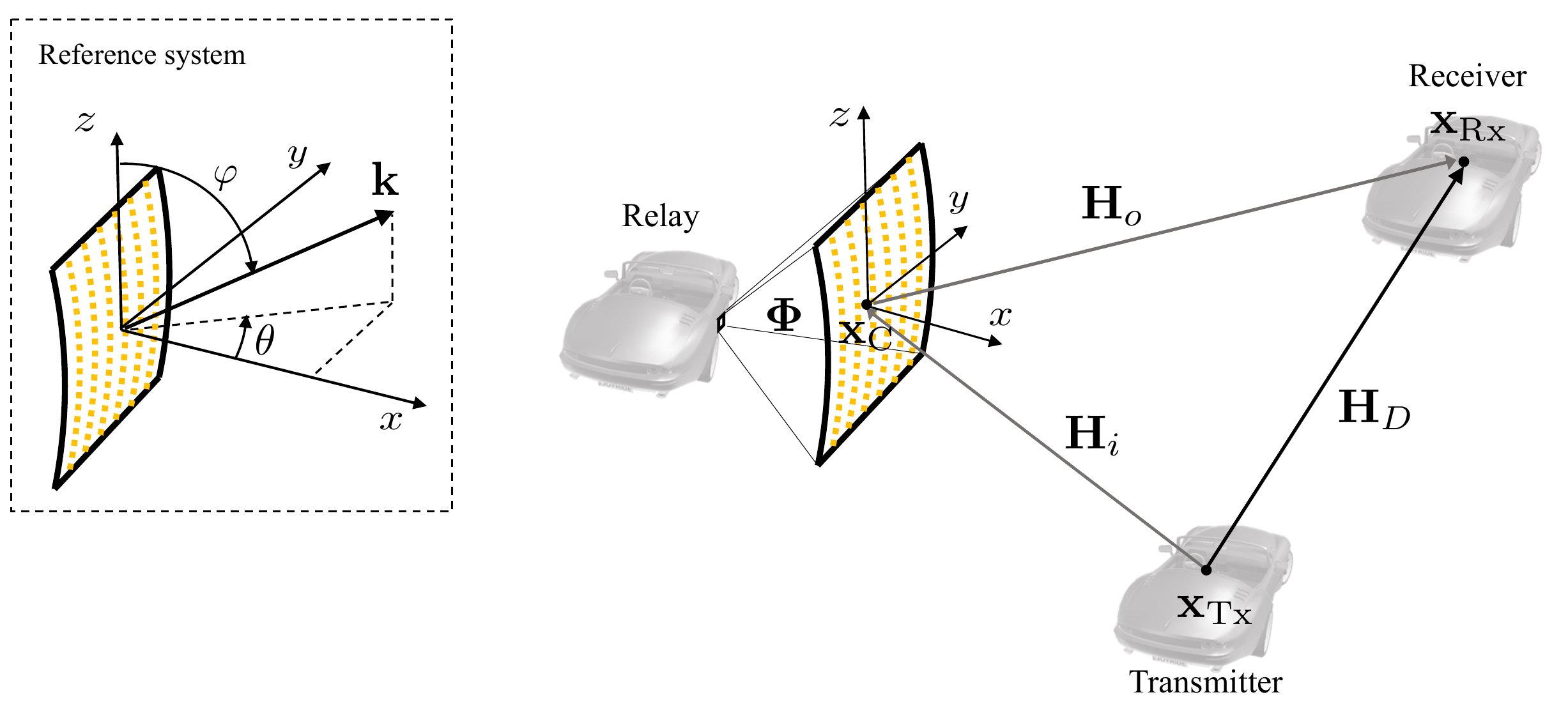}
    \caption{Sketch of the geometry of the considered system. Each vehicle is equipped with a \gls{CEMS} on doors to serve as an opportunistic relay.}
    \label{fig:systemModel}
\end{figure*}

The setup is the same as \cite{CIRS_TWC}, depicted in Figs. \ref{fig:MultiLaneScenario} and \ref{fig:systemModel}, where two vehicles (Tx and Rx) aim at establishing a V2V link with the aid of a relaying vehicle. The carrier frequency is $f_0$, and both Tx and Rx employ uniform linear arrays (ULAs) with $N_{\mathrm{Tx}}$ and $N_{\mathrm{Rx}}$ antennas. The relaying vehicle is equipped with \gls{CEMS}s on doors, each made by $M$ elements along the conformal coordinate (height, i.e., elevation $xz$ plane) and $N$ along the longitudinal one (length, i.e., azimuth $xy$ plane), for a total of $L=MN$ elements. The global coordinate system is such that the 3D position of Tx, Rx and relaying vehicle are $\mathbf{x}_{\mathrm{Tx}}$,  and $\mathbf{x}_{\mathrm{Rx}}$ and $\mathbf{x}_{\mathrm{C}}$. The absolute location of Tx-Rx arrays and the \gls{CEMS} identifies one reference element, e.g., the centroid. Cars move along direction $y$, the cross-motion axis is $x$ and $z$ denotes the vertical direction.

The spatial phase pattern across the \gls{CEMS} is pre-configured during the manufacturing process to \textit{(i)} compensate for the arbitrary shape of the car door, making it behave as a perfect \textit{flat} anomalous mirror for the incident signal and \textit{(ii)} obtaining an opportunistic low-cost relay to enable V2V communication in case the Tx-Rx link is blocked. The car door is modelled as a truncated cylinder of radius $R$. The position of the $mn$th element of the \gls{CEMS} in local coordinates (i.e., w.r.t. the \gls{CEMS} center) is  
\begin{equation}\label{eq:local_SPCEMS_element}
    \mathbf{p}_{mn} = \left[R (\cos \psi_m - 1), \, d_n (n-1),\, R \sin\psi_m \right]^T
\end{equation}
for $m = -M/2, \dots, M/2-1$ and $n = -N/2, \dots, N/2-1$, where $\psi_m = 2m\arcsin\left(\frac{d_m}{2R}\right)$ is the angular position in cylindrical coordinates of the $m$-th row of \gls{CEMS} elements, and $d_m$ and $d_n$ are the elements' spacing along the vertical and horizontal directions, respectively. The position of the $mn$th element in the global reference system is $\mathbf{x}_{mn} = \mathbf{x}_\mathrm{C}+ \mathbf{p}_{mn}$. The choice $R\in[1,8]$ m provides a curvature that is in line with common curved car doors \cite{DoorRadius}, and thus these reference values are used throughout the paper. Note that, the cylindrical shape is chosen to provide closed form phase patterns for the design methodologies outlined in Section \ref{sec:modular_design}. In any case, the latter optimization methods are general and they can be applied to any conformal shape of \gls{CEMS}.
%
\subsection{Signal and Channel Model}\label{subsect:signal_model}

The received signal at the Rx array can be written as follows
\begin{equation}\label{eq:receivedSignal_RIS}
    y = \mathbf{w}^H (\mathbf{H}_D + \mathbf{H}_R) \mathbf{f} \,s + \mathbf{w}^H\mathbf{n},
\end{equation}
where \textit{(i)} $s \in \mathbb{C} \sim \mathcal{CN}\left(0, \sigma_s^2\right)$ is the Tx information symbol; \textit{(ii)} $\mathbf{f} \in \mathbb{C}^{N_\mathrm{Tx} \times 1}$ is the spatial precoding vector at Tx side, such that $\| \mathbf{f}\|_F^2=N_\mathrm{Tx}$; \textit{(iii)} $\mathbf{H}_D \in \mathbb{C}^{N_\mathrm{Rx} \times N_\mathrm{Tx}}$ is the direct Tx-Rx channel; \textit{(iv)} $\mathbf{H}_R \in \mathbb{C}^{N_\mathrm{Rx} \times N_\mathrm{Tx}}$ is the relayed channel through the \gls{CEMS}; \textit{(v)} $\mathbf{w} \in \mathbb{C}^{N_\mathrm{Rx}\times 1}$ is the Rx combiner, such that $\| \mathbf{w}\|_F^2=N_\mathrm{Rx}$, while \textit{(vi)} $\mathbf{n} \sim \mathcal{CN}(\mathbf{0}, \sigma^2_n\mathbf{I}_{N_\mathrm{Rx}})$ is the additive noise at the Rx antennas, uncorrelated in space. Furthermore, we define $\zeta = \sigma^2_s / \sigma^2_n$ as the transmit signal-to-noise ratio (SNR).
%
The direct channel between Tx and Rx can be written under the \gls{FF} assumption, that typically holds for limited array size (few to tens of elements at high frequencies) and few to tens of meters of distance. Therefore, channel $\mathbf{H}_D$ follows the geometric model proposed in \cite{rappaport2019wireless}:
\begin{equation}\label{eq:directChannel}
    \mathbf{H}_D = \alpha\, g_\mathrm{Rx}({\xi}_\mathrm{Rx}) g_\mathrm{Tx}({\xi}_\mathrm{Tx}) \mathbf{a}_\mathrm{Rx}({\xi}_\mathrm{Rx})\mathbf{a}^H_\mathrm{Tx}({\xi}_\mathrm{Tx}), 
\end{equation}
where $\alpha$ is the complex scattering amplitude; $g_\mathrm{Tx}, ({\xi}_\mathrm{Tx})$ and $g_\mathrm{Rx}({\xi}_\mathrm{Rx})$ are the single antenna responses \cite{3GPP_AntennaPattern}, dependant on the pointing angles ${\xi}_\mathrm{Tx}$ and ${\xi}_\mathrm{Rx}$. $\mathbf{a}_\mathrm{Tx}({\xi}_\mathrm{Tx}) \in \mathbb{C}^{N_{Tx} \times 1}$ and $\mathbf{a}_\mathrm{Rx}({\xi}_\mathrm{Rx})  \in \mathbb{C}^{N_{Rx} \times 1}$ are the Tx and Rx steering vectors. We consider a single path as multiple paths do not affect the results herein. The amplitude $\alpha = PL_d^{-1/2}e^{j\rho}$, depends on the Geometric losses $PL_d$ \cite{CIRS_TWC}, that takes into account the free space path-loss, shadowing and blockage losses (see \cite{3GPPTR37885} for further details), and $\rho$ is assumed $\rho\sim\mathcal{U}[0, 2\pi)$, independent across different paths. %

The relayed link $\mathbf{H}_R$ through the \gls{CEMS} follows the general model \cite{CIRS_TWC}:
\begin{equation}\label{eq:reflectionChannel}
    \mathbf{H}_R =\mathbf{H}_{o}^H \boldsymbol{\Phi} \mathbf{H}_{i}
\end{equation}
where $\mathbf{H}_{i}\in\mathbb{C}^{L\times N_\mathrm{Tx}}$ is the forward channel from the Tx array to the \gls{CEMS}, $\boldsymbol{\Phi}$ is the phase configuration matrix of the \gls{CEMS}, herein assumed as diagonal $\mathbf{\Phi} = \mathrm{diag}\left(e^{j\phi_{1}},...,e^{j\phi_{L}}\right)$, and $\mathbf{H}_{o}\in\mathbb{C}^{L\times N_\mathrm{Rx}}$ is the backward channel from the \gls{CEMS} to the Rx array. While the \gls{FF} assumption generally holds for the direct Tx-Rx link, the link through the \gls{CEMS} can be either in \gls{FF} (with planar wavefront) or near-field (with curved wavefront) depending on the size of the \gls{CEMS} compared to the Tx/Rx-\gls{CEMS} distances. Thus, the incidence and reflection channel models, accounting for the propagation from the $\nu$th Tx antenna to the $\kappa$th Rx one, through the scattering from the $\ell=m\times n$ \gls{CEMS} element, are:
\begin{align}
    &\left[\mathbf{H}_{i}\right]_{\ell\nu} \hspace{-0.1cm}= \hspace{-0.1cm}[\mathbf{A}_i]_{\ell\nu}\, g_\mathrm{Tx}([\boldsymbol{\xi}_{\mathrm{Tx}}]_{\ell\nu}) \, g([\boldsymbol{\varTheta}_i,\boldsymbol{\varPhi}_i]_{\ell\nu})  e^{-j\frac{2\pi}{\lambda_0}[\boldsymbol{R}_i]_{\ell\nu}},\label{eq:forwardChannel}\\
    &\left[\mathbf{H}_{o}\right]_{\ell\kappa}\hspace{-0.1cm} = \hspace{-0.1cm}[\mathbf{A}_o]_{\ell\kappa}\,
    g([\boldsymbol{\varTheta}_o,\boldsymbol{\varPhi}_o]_{\ell\kappa})\,
    g_\mathrm{Rx}([\boldsymbol{\xi}_{\mathrm{Rx}}]_{\ell\kappa})  e^{-j\frac{2\pi}{\lambda_0}[\boldsymbol{R}_o]_{\ell\kappa}};\label{eq:backwardChannel}
\end{align}
$\boldsymbol{\xi}_{\mathrm{Tx}}$ and $\boldsymbol{\xi}_{\mathrm{Rx}}$ are the matrices of pointing and reception azimuth angles from each antenna of the Tx array to each element of the \gls{CEMS} and vice-versa from the each element of the \gls{CEMS} to each antenna of the Rx array, respectively; $g(\cdot)$ is the single-element gain of the \gls{CEMS} \cite{RadPatt}; $\boldsymbol{\varTheta}_i$ and  $\boldsymbol{\varPhi}_i$ are the matrices of azimuth and elevation angles of incidence (AoI) onto the \gls{CEMS} from each element of the Tx array, and a similar definition applies to the angles of reflection (AoR) $\boldsymbol{\varTheta}_o$ and $\boldsymbol{\varPhi}_o$; the scattering amplitudes of the incidence and reflection channels are:
\begin{equation}
    [\mathbf{A}_i]_{\ell \nu} \hspace{-0.1cm}= \hspace{-0.1cm} \frac{\lambda_0}{4 \pi [\boldsymbol{R}_i]_{\ell \nu}},\,\, [\mathbf{A}_o]_{\ell \kappa} \hspace{-0.1cm}= \hspace{-0.1cm} \frac{\lambda_0}{4 \pi [\boldsymbol{R}_o]_{\ell\kappa}};
\end{equation}
$\boldsymbol{R}_i$ and $\boldsymbol{R}_o$ denote the matrices of Tx-\gls{CEMS} and \gls{CEMS}-Rx distances, such that:
\begin{equation*}
    [\boldsymbol{R}_i]_{\ell \nu} \hspace{-0.1cm}= \hspace{-0.1cm} \| \mathbf{x}_{\ell} \hspace{-0.1cm}- \hspace{-0.1cm} \mathbf{x}_{\mathrm{Tx},\nu} \|_F,\,\,\,\,\, [\boldsymbol{R}_o]_{\ell\kappa} \hspace{-0.1cm}= \hspace{-0.1cm}  \| \mathbf{x}_{\mathrm{Rx},\kappa}  \hspace{-0.1cm}- \hspace{-0.1cm}\mathbf{x}_{\ell}\|_F.
\end{equation*}

\section{Phase Configuration for CEMS}\label{sect:SPCEMS_config}

Models \eqref{eq:forwardChannel} and \eqref{eq:backwardChannel} account for the possible near-field operating condition of the relayed link, due to the size of the CEMS that we assume larger than Tx and Rx arrays. Phase configuration of the CEMS in near-field would require the double \textit{focusing} of the scattered signal from the position of the Tx array to the position of the Rx to maximize the received signal power \cite{NearFieldEmil}. However, in a dynamic scenario such as the V2V one, the relative Tx-CEMS-Rx distance is rapidly varying, thus any mismatch with the phase configuration of the CEMS would lead to drastic power losses at the Rx side, with consequent outages. The latter aspect might be addressed with the use of reconfigurable CEMS, but the high implementation costs (as well as the required signaling for practical operation, e.g., to estimate $\boldsymbol{R}_i$ and $\boldsymbol{R}_o$) call for simpler design approaches for passive pre-configured CEMS. In this setting, we opt for a simpler \gls{FF} phase design approach, i.e., matrix $\mathbf{\Phi}$ is designed according to one or multiple pairs of AoI/AoR, namely $\boldsymbol{\vartheta}_{i} = (\theta_{i},\varphi_{i})$, $\boldsymbol{\vartheta}_{o} = (\theta_{o},\varphi_{o})$, 
defined by the centers of Tx/Rx and CEMS. In other words, the propagation between Tx/Rx and CEMS can be either near-field or \gls{FF}, but the phase design of the CEMS follows the \gls{FF} assumption.

For an arbitrarily shaped CEMS, the generalized Snell's law dictates the the phase to be applied at each element to manipulate the reflection properties. Let us define the incident and reflected wavevectors: 
\begin{align}
    \mathbf{k}_{i}(\boldsymbol{\vartheta}_i) & = \frac{2 \pi}{\lambda_0} \left[\sin \varphi_{i} \cos\theta_{i}, \sin \varphi_{i} \sin\theta_{i}, \cos\theta_i\right]^T\label{eq:inc_K}\\
    \mathbf{k}_{o}(\boldsymbol{\vartheta}_o) & = \frac{2 \pi}{\lambda_0} \left[\sin \varphi_{o} \cos\theta_{o}, \sin \varphi_{o} \sin\theta_{o}, \cos\theta_o\right]^T,\label{eq:out_K}
\end{align}
The optimal phase at the $(m,n)$th element of the CEMS to enable the anomalous reflection from $\boldsymbol{\vartheta}_i$ to $\boldsymbol{\vartheta}_o$ is 
\begin{equation}\label{eq:phase_config}
\begin{split}
     \phi_{\ell}(\boldsymbol{\vartheta})& = \phi_{mn}(\boldsymbol{\vartheta}) = \mathbf{p}^T_{mn}\mathbf{k}_i(\boldsymbol{\vartheta}_i)  + \mathbf{p}_{mn}^T\mathbf{k}_o(\boldsymbol{\vartheta}_o),
\end{split}
\end{equation}
where $\boldsymbol{\vartheta} = (\boldsymbol{\vartheta}_i,\boldsymbol{\vartheta}_o)$ is the joint set of AoI and AoR, and $\mathbf{p}_{mn} = [x_{mn},y_{mn},z_{mn}]^T$ of the $(m,n)$th is the position of the element of the CEMS. The optimal configuration of the CEMS would require the statistical distribution of both azimuth \textit{and} elevation AoI and AoR, which is scenario-dependent and not easy to be obtained, in general. We can reduce the design complexity by leveraging the geometry of the considered V2V scenario, where all the vehicles have similar heights, implying that elevation AoI and AoR tend to $\varphi_i=\varphi_o=\pi/2$ (Fig. \ref{fig:systemModel}). Note that this assumption holds when the distance among the vehicles is much larger than the height difference (i.e., the difference between the vertical coordinate of the arrays/CEMS). If the difference between the elevation AoI/AoR to/from the CEMS is less than the reflection beamwidth (that can be evaluated as in \cite{CIRS_TWC}), the approximation $\varphi_i=\varphi_o=\pi/2$ is acceptable. The latter simplifies the design of the CEMS. \\ By plugging \eqref{eq:local_SPCEMS_element} and $\varphi_i=\varphi_o=\pi/2$ into \eqref{eq:phase_config}, we obtain the phase configuration for cylindrical EMS \cite{CIRS_TWC}, reported in \eqref{eq:phase_config_FF_azimuth},
\begin{figure*}[!t]
\begin{equation}\label{eq:phase_config_FF_azimuth}
      \phi_{mn}(\theta_i,\theta_o) = \frac{4 \pi}{\lambda_0} \cos\left(\frac{\theta_i-\theta_o}{2} \right) \left[R (\cos\psi_m -1) \cos\left(\frac{\theta_i+\theta_o}{2} \right) + d_n (n-1) \sin\left(\frac{\theta_i+\theta_o}{2} \right)\right]
\end{equation}
\hrulefill
\end{figure*}
depending solely on azimuth angles $\theta_i$ and $\theta_o$. As expected, the phase gradient is linear along the cylindrical coordinate (index $n$) but it is non-linear along the curved coordinate (index $m$), as function of the curvature radius $R$. Now, the CEMS can be designed according to the distribution of $\theta_i$ and $\theta_o$. We remark that, although the paper considers the CEMS design along azimuth \gls{AoI} and \gls{AoR}, the contribution here can be extended to the elevation whenever a joint distribution of both angles is available. 

By plugging a specific phase pattern in either \eqref{eq:phase_config} or \eqref{eq:phase_config_FF_azimuth}, function of a desired configuration angle pair $\overline{\boldsymbol{\vartheta}} = (\overline{\boldsymbol{\vartheta}}_i, \overline{\boldsymbol{\vartheta}}_o)$, the reflection channel becomes:
\begin{equation}\label{eq:dynamic_channel}
\begin{split}
     \left[\mathbf{H}_{R}\right]_{\kappa \nu} & =  \sum_{\ell}g_{\kappa \nu \ell} {A}_{\kappa \nu \ell} e^{-j\frac{2\pi}{\lambda_0}\left([\boldsymbol{R}_i]_{\ell\nu} +[\boldsymbol{R}_o]_{\ell\kappa} \right)} e^{j\phi_{\ell}(\overline{\boldsymbol{\vartheta}})} \\
     & \overset{(\mathrm{FF}) }{\approx} g \widetilde{A}_{\kappa \nu} \sum_{\ell} e^{j (\phi_{\ell}(\overline{\boldsymbol{\vartheta}}) - \phi_{\ell}(\boldsymbol{\vartheta})) } 
\end{split}
\end{equation}
where $g_{\kappa \nu \ell}$ is the product of all the single-element responses detailed in \eqref{eq:forwardChannel}-\eqref{eq:backwardChannel}, $\mathbf{A}_{\kappa \nu \ell} = [\mathbf{A}_i]_{\ell\nu}\; [\mathbf{A}_o]_{\ell\kappa}$ is the contribution of the amplitudes. The approximation is for \gls{FF} operation, thus the propagation phase can be decomposed as follows
\begin{equation}\label{eq:FF_propagation_phase}
    \frac{2\pi}{\lambda_0}\left([\boldsymbol{R}_i]_{\ell\nu} \hspace{-0.1cm}+\hspace{-0.1cm}[\boldsymbol{R}_o]_{\ell\kappa} \right)  \overset{(\mathrm{FF}) }{\approx} \frac{2\pi}{\lambda_0} (r_i + r_o) + \phi_\kappa + \phi_\nu + \phi_{\ell}(\boldsymbol{\vartheta})
\end{equation}
where the first term is the propagation phase between the center of the arrays and the center of the CEMS, $r_i = \| \mathbf{x}_\mathrm{C} - \mathbf{x}_\mathrm{Tx} \|$, $r_o = \| \mathbf{x}_\mathrm{Rx} - \mathbf{x}_\mathrm{C} \|$, the second and third terms are the phases at Tx and Rx (to be compensated by precoding and combining) and the last term is the true propagation phase across the CEMS, function of the true AoI and AoR $\boldsymbol{\vartheta}$. The first three terms in \eqref{eq:FF_propagation_phase} are included in the constant amplitude term $\widetilde{A}_{\kappa \nu}$, while $g_{\kappa \nu \ell} \rightarrow g$. We notice that if the phase shifts of the EM is designed according to \gls{FF} assumption, the propagation phase under near-field operation cannot be perfectly compensated, although the advantages compared to a bare car door (without the CEMS) in terms of SNR and spectral efficiency are remarkable. In \gls{FF}, instead, the reflection channel gain $\|\mathbf{H}_R\|_F^2$ is proportional to the phase mismatch between the true AoI/AoR $\boldsymbol{\vartheta}$ and the imposed ones $\overline{\boldsymbol{\vartheta}}$. 

For static and pre-configured CEMS, choosing a single AoI-AoR pair allows using a relaying vehicle for a single specific angular direction connecting Tx and Rx. This is the contribution of our previous work \cite{CIRS_TWC}, that is limited to specular reflection. This work aims at designing the CEMS as an anomalous mirror allows covering an entire set of AoI/AoR, exploiting traffic statistics and the distribution of AoI/AoR. In the following, we detail two CEMS design approaches aimed at maximizing  either the spectral efficiency or the coverage probability. Two methods are proposed: \textit{(i)} optimization of the whole CEMS, whereby the phase of all the elements is independently designed according to the continuous AoI/AoR distribution and \textit{(ii)} optimization of the phase of the CEMS, whereby the latter is conveniently split in sub-regions (called \textit{modules}) and the AoI/AoR domain is discretized. The former method provides a phase patterm that is not constrained to the form of either \eqref{eq:phase_config} or \eqref{eq:phase_config_FF_azimuth}, but rather it provides an arbitrary-shaped non-linear phase gradient. The second method for modular CEMS, instead, provides a distinct phase pattern for each module, whose form is according \eqref{eq:phase_config_FF_azimuth}. 

\section{Unstructured CEMS Phase Optimization}\label{sec:Opt_Unstructured}

The statistical characterization of AoI and AoR play a pivotal role in characterizing the scenario, thereby facilitating the probabilistic formulation of the average \gls{SE} and coverage maximization. In the following, we formulate the \textit{unstructured} phase design problem in both cases, meaning that the phase of the CEMS is not constrained to comply with neither \eqref{eq:phase_config} nor \eqref{eq:phase_config_FF_azimuth}. The computational complexity of this general design approach is then discussed and motivates the structured approach in Section \ref{sec:modular_design}.

\subsection{Spectral Efficiency Optimization}

The first design option for the phase of the CEMS involves the maximization of the average spectral efficiency over multi-lane vehicular environment. Let us denote with $\mathbf{X} = [\mathbf{x}_\mathrm{Tx}, \mathbf{x}_\mathrm{Rx}, \mathbf{x}_\mathrm{C}]\in \mathbb{R}^{3 \times 3}$ the matrix of the joint position of Tx, Rx and CEMS in space. We assume to have the a-priori distribution of $\mathbf{X}$ that we denote with $f(\mathbf{X})$, whose domain is $\mathcal{X}$. The distribution $f(\mathbf{X})$ can be mapped into the distribution of $\theta_i$ and $\theta_o$ by means of proper transformations. Note that the distribution $f(\mathbf{X})$ used in this paper for the CEMS design can in practice be the distribution of any other parameter of interest affecting the objective function, of which we have an a-priori knowledge. The phase design follows from the solution of the following weighted maximization:
\begin{subequations}\label{eq:SNR_opt}
\begin{alignat}{2}
&\underset{\boldsymbol{\Phi}\,\,}{\mathrm{max}}      &\quad& \sum_{{\mathbf{X} \in \mathcal{X}}} f(\mathbf{X}) \,\mathrm{log}_2\left(1+\zeta\left\|\mathbf{h}^H_{o}(\mathbf{X})\,\boldsymbol{\Phi}\,\mathbf{h}_{i}(\mathbf{X})\,\right\|_F^2\right) 
\\
&\mathrm{s.t.} &      & 0 \leq \phi_{\ell} < 2\pi, \,\,\,\,  \ell = 1,\cdots, L,
\end{alignat}
\end{subequations}
where $\mathbf{h}_{o}(\mathbf{X}) = \mathbf{H}_o \mathbf{w}\in \mathbb{C}^{L\times 1}$ and $\mathbf{h}_{i}(\mathbf{X}) =  \mathbf{H}_i\mathbf{f} \in \mathbb{C}^{L\times 1}$ are the combined and precoded versions of $\mathbf{H}_o$ and $\mathbf{H}_i$, assuming that Tx and Rx know the location of the relay. 
The problem expressed in \eqref{eq:SNR_opt} can be reformulated in accordance with \cite{OPT_SNR_1}, through the subsequent mathematical manipulation
\begin{subequations}\label{eq:sdr_SNRopt}
\begin{alignat}{2}
&\underset{\mathbf{V}\,\,}{\mathrm{max}}      &\quad&  \sum_{{\mathbf{X} \in \mathcal{X}}} 
\,\mathrm{log}_2(1+\zeta \, \mathrm{tr}({\mathbf{R}}(\mathbf{X})\mathbf{V}))\\
&\mathrm{s.t.} &      & \mathbf{V}_{\ell,\ell} = 1 \,\,\,\,  \ell = 1,\cdots, L, \\
& &      & \mathbf{V} \succeq 0,\\
& &      & \mathrm{rank}(\mathbf{V}) = 1 \label{eq:rank1}, 
\end{alignat}
\end{subequations}
where $\mathbf{R}(\mathbf{X}) = \left(\mathrm{diag}(\mathbf{h}_o) \mathbf{h}_i \mathbf{h}^H_i \mathrm{diag}(\mathbf{h}^*_o)\right) f(\mathbf{X})$ and 
$\mathbf{V} = \mathbf{v} \mathbf{v}^{\mathrm{H}} \in \mathbb{C}^{L \times L}$ where $\mathbf{v} = [e^{j\phi_1}, \hdots,e^{j\phi_L}]^T$. Now, the optimization is carried out over $\mathbf{V}$, that is positive semi definite with rank 1. Optimization problem \eqref{eq:sdr_SNRopt} is non-convex due to the presence of the rank-1 constraint. Several techniques have been proposed to tackle the problem, e.g., see  \cite{Zhang_Randomization}. The latter relaxes the problem by discarding the rank-1 constraint and obtaining a higher-rank solution that is then refined by further heuristic search. The method in \cite{Zhang_Randomization} is known to provide an efficient solution (in terms of computational complexity) that is however non-optimal. 

To address problem \eqref{eq:sdr_SNRopt}, we define a  penalty-based algorithm, where the non-convex rank-1 constraint in \eqref{eq:rank1} is expressed equivalently using the following equality (see \cite{OPT_RATE_3}) 
\begin{equation}
   \left\|\mathbf{V}\right\|_{*} - 
   \left\| \mathbf{V}\right\|_{2} = \mathrm{tr}(\mathbf{\Lambda}_\mathbf{V}) - \mathrm{max}(\mathbf{\Lambda}_\mathbf{V})  = 0
\end{equation}
where $\left\|\cdot\right\|_{*}$ and $\left\|\cdot\right\|_{2}$ represents the nuclear and spectral norms, respectively the sum and the maximum of the eigenvalues of $\mathbf{V}$ (collected in matrix $\mathbf{\Lambda}_\mathbf{V}$). For $ \mathbf{V} \in \mathbb{H}^{L\times L}$, $\left\|\mathbf{V}\right\|_{*} - \left\| \mathbf{V}\right\|_{2}\geq 0$ and the above equality is satisfied only if the matrix $\mathbf{V}$ exhibits rank 1. Therefore, we reformulate the problem in \eqref{eq:sdr_SNRopt} as
\begin{subequations}
\begin{alignat}{2}\label{eq:PenaltyModel}
&\underset{\mathbf{V}\,\,}{\mathrm{max}}      &\quad&  \sum_{{\mathbf{X} \in \mathcal{X}}} 
\mathrm{log}_2\left(1\hspace{-0.1cm}+\hspace{-0.1cm}\zeta \, \mathrm{tr}({\mathbf{R}}(\mathbf{X})\mathbf{V})\right)\hspace{-0.1cm}-\hspace{-0.1cm} \eta \left(\left\|\mathbf{V}\right\|_{*} \hspace{-0.1cm}- \hspace{-0.1cm}
   \left\| \mathbf{V}\right\|_{2}\right)\\
&\mathrm{s.t.} &      & \mathbf{V}_{\ell,\ell} = 1 \,\,\,\,  \ell = 1,\hdots, L, \\
& &      & \mathbf{V} \succeq 0.  
\end{alignat}
\end{subequations}
where the term $\eta \in \mathbb{R}^{+}$ represents the penalty term. The objective function \eqref{eq:PenaltyModel} results in a balance between rank minimization and average SE maximization.  

Problem \eqref{eq:PenaltyModel} is still non-convex due to nuclear and spectral norms. Rearranging \eqref{eq:PenaltyModel}, one obtains an iterative cone program as in \cite{OPT_RATE_3}, that at the $i$th iteration is as follows: 
\begin{subequations}
\begin{alignat}{2}\label{eq:PenaltyModelSNR}
&\underset{\mathbf{V}, \mathbf{Z}, \mathbf{t}\,\,}{\mathrm{max}}      &\quad&  \sum_{\mathbf{X} \in \mathcal{X}} t(\mathbf{X})- \eta \left(\mathrm{tr}(\mathbf{V}^{\mathrm{T}} \mathbf{Z}) - \delta(\mathbf{V}^{(i)},\mathbf{V})\right)\\
&\mathrm{s.t.} &      & \mathbf{V}_{\ell,\ell} = 1 \,\,\,\,  \ell = 1,\hdots, L, \\
& &      & \mathbf{V} \succeq 0\label{re:sdp},\\
& &      & 	\begin{bmatrix}
\mathbf{I} & \mathbf{Z}^{\mathrm{T}}\\
\mathbf{Z} & \mathbf{I}
\end{bmatrix} \succeq 0\label{re:nuclearsdp}\\
& &     & t(\mathbf{X}) \leq
\,\mathrm{log}_2(1+\zeta \, \mathrm{tr}({\mathbf{R}}(\mathbf{X})\mathbf{V})), \forall {\mathbf{X}}.\label{re:expcone}
\end{alignat}
\end{subequations}
Here, $\mathbf{t} = [t(\mathbf{X})]_{\mathbf{X} \in \mathcal{X}}$ is the vector of auxiliary variables upper-bounding the spectral efficiency. The nuclear norm is rewritten as $\left|\left|\mathbf{V}\right|\right|_{*} = \mathrm{sup}_{\left|\left|\mathbf{Z}\right|\right|_2 \leq 1} \mathrm{tr}(\mathbf{V}^{\mathrm{T}} \mathbf{Z})$, by means of an auxiliary variable $\mathbf{Z} \in \mathbb{C}^{L \times L}$, while  the spectral norm is linearized by first-order Taylor expansion at the $i$th iteration as
\begin{align} 
\begin{split}
   \left|\left| \mathbf{V}\right|\right|_{2} &\geq  \left|\left| \mathbf{V}\right|\right|_{2}^{(i)} + \mathrm{tr}\left(\mathbf{u}(\mathbf{V}^{(i)}) \mathbf{u}(\mathbf{V}^{(i)})^{\mathrm{H}}(\mathbf{V}-\mathbf{V}^{(i)})\right)\\
   &\geq\delta(\mathbf{V}^{(i)},\mathbf{V})
\end{split}
\end{align}
where $\mathbf{u}(\mathbf{V}^{(i)})$ represents the eigenvector corresponding to the largest eigenvalue of $\mathbf{V}^{(i)}$ (solution at the $i$th iteration). Now, we attain an optimal solution for the original non-convex problem posited in \eqref{eq:sdr_SNRopt} by iteratively solving \eqref{eq:PenaltyModelSNR} using MOSEK software tool with a judicious tuning of the parameter~$\eta$.

\subsection{Coverage Optimization}\label{subsect:unstructured_coverage}
Coverage optimization aims at maximizing the coverage probability over the area of interest, building upon the SDR program in \eqref{eq:PenaltyModelSNR}. Therefore, we can set the CEMS phase design according to coverage maximization as follows:
\eqref{eq:PenaltyModelSNR} as
\begin{subequations}
\begin{alignat}{2}\label{eq:PenaltyModelSNR2}
&\underset{\mathbf{V}, \mathbf{Z}, \mathbf{b}}{\mathrm{max}}      &\qquad&  \sum_{{\mathbf{X} \in \mathcal{X}}} b(\mathbf{X})
- \eta \left(\mathrm{tr}(\mathbf{V}^{\mathrm{T}} \mathbf{Z}) - \delta(\mathbf{V}^{(n)},\mathbf{V})\right)\\
&\mathrm{s.t.} &      & \mathbf{V}_{\ell,\ell} = 1 \,\,\,\,  \ell = 1,\hdots, L, \\
& &      & \mathbf{V} \succeq 0,\\
& &      & 	\begin{bmatrix}
\mathbf{I} & \mathbf{Z}^{\mathrm{T}}\\
\mathbf{Z} & \mathbf{I}
\end{bmatrix} \succeq 0\\
&  &      &\zeta \left\|\mathbf{h}^H_{o}(\mathbf{X})\boldsymbol{\Phi}\mathbf{h}_{i}(\mathbf{X})\,\right\|^2 \geq \gamma\,b(\mathbf{X}), \,\,\,\forall {\mathbf{X} }
\end{alignat}
\end{subequations}
where
\begin{align}\label{eq:coverage_bool}
b(\mathbf{X})   = 
\begin{cases}
1 & \text{if } \zeta \left\|\mathbf{h}_{o}(\mathbf{X})^H\,\boldsymbol{\Phi}\,\mathbf{h}_{i}(\mathbf{X})\,\right\|_F^2  > \gamma \,\\
0 & \text{otherwise},
\end{cases}
\end{align}
is a binary variable indicating whether a certain configuration $\mathbf{X}$ provides coverage or not (SNR above a pre-defined threshold). The set of binary variables \eqref{eq:coverage_bool} is stacked in vector $\mathbf{b} = [b(\mathbf{X})]_{\mathbf{X}\in\mathcal{X}}$. However, \eqref{eq:PenaltyModelSNR2} is in this case a mixed-integer semi-definite program (SDP), that is known to be difficult to solve from computational complexity point of view \cite{gally2018framework}. Therefore, we need to reformulate the problem by eliminating binary variables. We can write  
\begin{subequations}
\begin{alignat}{2}\label{eq:PenaltyModelSNR3}
&\underset{\mathbf{V}, \mathbf{Z}, \mathbf{y}, \mathbf{Y}\,\,}{\mathrm{max}}      &\quad&  \sum_{{\mathbf{X} \in \mathcal{X}}} y(\mathbf{X})- \eta \left(\mathrm{tr}(\mathbf{V}^{\mathrm{T}} \mathbf{Z}) - \delta(\mathbf{V}^{(n)},\mathbf{V})\right)\\
&\mathrm{s.t.} &      & \mathbf{V}_{\ell,\ell} = 1 \,\,\,\,  \ell = 1,\hdots, L, \\
& &      & \mathbf{V} \succeq 0\label{eq:sdp1},\\
& &      & 	\begin{bmatrix}
\mathbf{I} & \mathbf{Z}^T\\
\mathbf{Z} & \mathbf{I}
\end{bmatrix} \succeq 0\label{eq:sdp2},\\
&  &      &\zeta \left\|\mathbf{h}^H_{o}(\mathbf{X})\boldsymbol{\Phi}\mathbf{h}_{i}(\mathbf{X})\right\|_F^2 \geq \gamma\, y(\mathbf{X}), \forall {\mathbf{X} },\\
&  & & \begin{bmatrix}
1 & \mathbf{y}^T\\
\mathbf{y} & \mathbf{Y}
\end{bmatrix} \succeq 0\label{eq:sdp3},\\
&  & & \mathrm{diag}(\mathbf{Y}) = \mathbf{y}.
\end{alignat}
\end{subequations}
where $\mathbf{y} = [y(\mathbf{X})]_{\mathbf{X}\in \mathcal{X}}\in \mathbb{R}^{|\mathcal{X}| \times 1}$ and $\mathbf{Y} \in \mathbb{R}^{|\mathcal{X}| \times |\mathcal{X}|}$ are new auxiliary variables such that $\mathrm{diag}(\mathbf{Y}) = \mathbf{y}$. This relaxation eliminates binary variables, resulting in a reduction in the problem computational complexity.

\subsection{Complexity Considerations}\label{subsect:unstructured_complexity}
Computational complexity for \gls{SE} \eqref{eq:PenaltyModelSNR} or coverage optimization \eqref{eq:PenaltyModelSNR3}, can be expressed as a function of the number of CEMS elements as well as the cardinality of $\mathcal{X}$. In particular, the iterative cone program \eqref{eq:PenaltyModelSNR} involves 2 semidefinite constraints in \eqref{re:sdp} and \eqref{re:nuclearsdp} of dimension $L$ and $2L$, respectively. Moreover, the problem includes the  exponential cone constraint \eqref{re:expcone} of dimension $|\mathcal{X}|$. In general, the complexity associated to semidefinite constraints is higher compared to exponential cone constraints. Therefore, obtaining the exact computational complexity of such a mixed cone program is not straightforward and out of the scope of the paper. 
The overall complexity of a problem with only semidefinite constraints (excluding \eqref{re:expcone}) follows from \cite{park2018semidefinite} as $\mathcal{O}(V^2 L^{2.5} \log(\frac{1}{\epsilon}))$, where $V$ represents the total number of decision variables, that in \eqref{eq:PenaltyModelSNR} and \eqref{eq:PenaltyModelSNR3} are greater than $L$. This complexity is required to achieve an $\epsilon$-optimal solution, ensuring that $\|\mathbf{V} - \mathbf{V}_{opt} \|_F \leq \epsilon$ \cite{park2018semidefinite}. On the other hand, obtaining an $\epsilon$-optimal solution of the cone program with $|\mathcal{X}|$ exponential cone constraints requires $\mathcal{O}(V^2|\mathcal{X}|^{1.5}\log(\frac{1}{\epsilon}))$ operations \cite{chen2023exponential}. The problems in \eqref{eq:PenaltyModelSNR} and \eqref{eq:PenaltyModelSNR3} are solved in a iterative way, therefore the resulting complexity has to be multiplied by the maximum number of iterations.

\section{Structured CEMS Phase Optimization for Modular Architecture}\label{sec:modular_design}


As detailed in the previous section, unstructured methods are powerful tools that allow design CEMS phase profile at each single element. However, the complexity of such unstructured phase design prevents scaling up with the CEMS size, as needed in the considered high-frequency application, where CEMS shall cover all doors and a wide portion of the car, (with hundreds or thousands of elements) to provide a sufficient reflection gain. Therefore, we propose an alternative  \textit{structured} phase design approach that consists of a dimensionality reduction for the phase design problem. The main idea of the proposed method is to design the phase profile of few portions of the CEMS (called \textit{modules}) according to a pair of AoI and AoR from a pre-defined codebook, then using \eqref{eq:phase_config_FF_azimuth} to relate the AoI-AoR with the CEMS elements phase. Now, the number of parameters to be estimated is $2 P$, i.e., two angles for each of the $P$ modules. Of course, the number of modules $P$ is a hyper-parameters to be determined through numerical evaluation.

In this setting, we assume that the CEMS is composed of $P$ modules along the cylindrical direction, i.e., each module is made by $L'=M \times \frac{N}{P}$ elements. Each module operates according to a pair of selected azimuth angles $\overline{\boldsymbol{\theta}} = (\overline{\theta}_i,\overline{\theta}_o)$, to be selected from a codebook $\mathcal{A}_{\overline{\boldsymbol{\theta}}} = \{\overline{\boldsymbol{\theta}}_{1}, \overline{\boldsymbol{\theta}}_{2}, \cdots, \overline{\boldsymbol{\theta}}_{N_A}\}$, with $N_A$ entries. One module out of the $P$ ones can be configured according to only one of the angle pairs of the codebook, allowing multiple, optimized anomalous reflections. Goal of the phase design is to select the best angle pairs for either spectral efficiency or coverage maximization. The result is herein conveniently expressed as a vector of indices within the codebook $\mathcal{I} = \{i_1, i_2, \cdots, i_P\}$, representing selections from $\mathcal{A}_{\overline{\boldsymbol{\theta}}}$. The overall search space is $\mathcal{S}$, comprising all the ${N_A}^P$ possibilities. With a modular architecture of the CEMS, and a given selection vector $\mathcal{I}$, the phase matrix is block diagonal as
\begin{equation}\label{eq:modular_phase_matrix}
    \mathbf{\Phi}_{\mathcal{I}} = \mathrm{blkdiag}\left(\mathbf{\Phi}_{i_1}, \mathbf{\Phi}_{i_2},...,\mathbf{\Phi}_{i_P} \right)
\end{equation}
where $\mathbf{\Phi}_{i_p} = \mathrm{diag}(e^{j\phi_{i_p,1}},...,e^{j\phi_{i_p,L'}}) \in \mathbb{C}^{L'\times L'}$, with straightforward meaning.

\subsection{Spectral Efficiency Optimization}\label{subsect:structured_SE}
The modular CEMS can be designed according to the average spectral efficiency maximization as follows
\begin{subequations}
\begin{alignat}{2}
& \underset{\mathcal{I}}{\mathrm{max}} &\quad& \sum_{\mathbf{X} \in \mathcal{X}} \log_2\left(1 + \zeta \left\| \mathbf{h}_o(\mathbf{X})^H \mathbf{\Phi}_{\mathcal{I}} \mathbf{h}_i(\mathbf{X}) \right\|^2\right) f(\mathbf{X}) \label{eq:SEOptimizationObjective}\\
& \mathrm{s.t.} &      & i_p \in \{1, 2, \ldots, N_A\}, \quad \forall p = 1, 2, \ldots, P, \label{eq:IndexConstraint}\\
& & & |\mathcal{I}| = P\label{eq:CardinalityConstraint}
\end{alignat}
\end{subequations}
where $\mathbf{\Phi}_{\mathcal{I}}$ is for \eqref{eq:modular_phase_matrix}, configured for a given set of angles identified by $\mathcal{I}$.


\subsection{Coverage Optimization}\label{subsect:structured_coverage}

\begin{figure}[t!]
    \centering
    \begin{minipage}{0.48\textwidth}
       \centering
   \includegraphics[width=\linewidth]{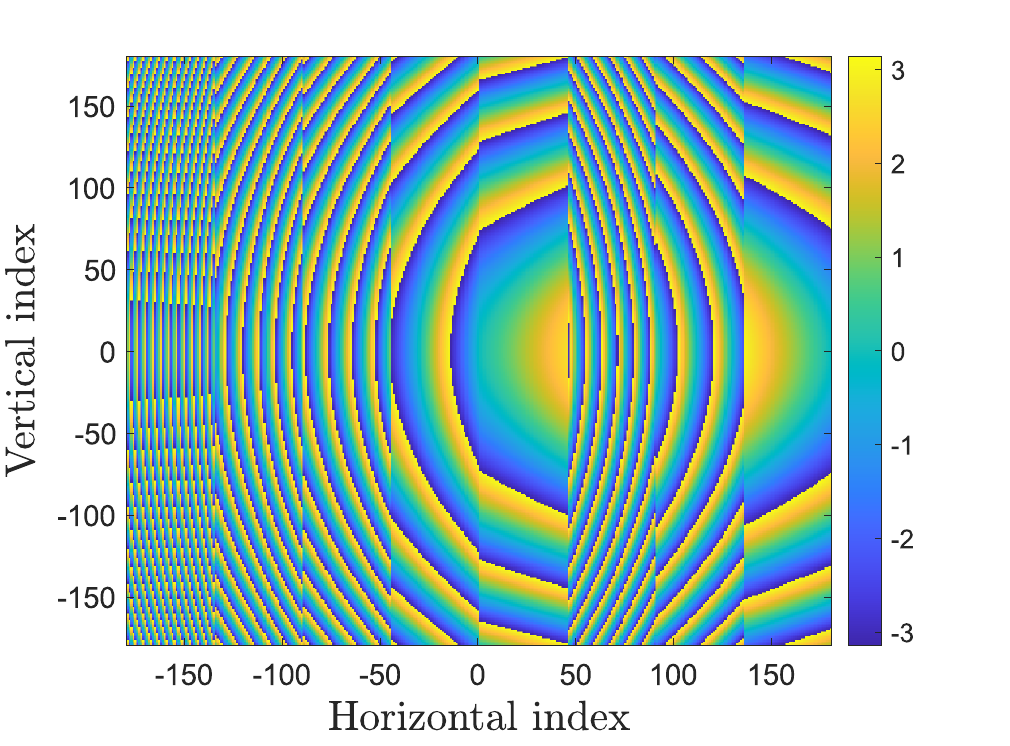}        \caption{Exemplary phase pattern for a $360\times 360$ CEMS made by $P=8$ modules, obtained with the structured modular design (either \eqref{eq:SEOptimizationObjective} or \eqref{eq:CoverageOptimizationObjective}).}
   \label{fig:PS8}
   \end{minipage}
\end{figure}

Pairwise with Section \ref{subsect:unstructured_coverage}, we can also design the modular CEMS for coverage maximization: 
\begin{subequations}
\begin{alignat}{2}
& \underset{\mathcal{I}}{\mathrm{max}} &\quad& \sum_{\mathbf{X} \in \mathcal{X}} b(\mathbf{X}, \mathcal{I})\, f(\mathbf{X}) \label{eq:CoverageOptimizationObjective}\\
& \mathrm{s.t.} &      & i_p \in \{1, 2, \ldots, N_A\}, \quad \forall p = 1, 2, \ldots, P, \label{eq:CoverageIndexConstraint}\\
& & & |\mathcal{I}| = P, \label{eq:CoverageCardinalityConstraint}
\end{alignat}
\end{subequations}
where $b(\mathbf{X}, \mathcal{I})$ is defined, similarly to \eqref{eq:coverage_bool}, as
\begin{align}
b(\mathbf{X}, \mathcal{I}) = 
\begin{cases}
1 & \text{if } \zeta \, \left\|\mathbf{h}_{o}^H(\mathbf{X})\,\mathbf{\Phi}_{\mathcal{I}}\,\mathbf{h}_{i}(\mathbf{X})\,\right\|^2 > \eta, \\
0 & \text{otherwise}.
\end{cases}
\end{align}
It is worth remarking that the optimization of a multi-module CEMS ($P>1$) cannot be addressed module-wise. Indeed, the SE and coverage performance is contingent upon the selection of the configuration of \textit{all} modules, as certain configurations that exhibit favorable performance for single modules may lead to detrimental interference when combined with other modules. An example of phase pattern obtained with either \eqref{eq:SEOptimizationObjective} or \eqref{eq:CoverageOptimizationObjective} is shown in Fig. \ref{fig:PS8}, for a $360 \times 360$ \gls{CEMS} made by $P=8$ modules. The horizontal (azimuth) phase gradient is different for each module, as expected, but  a vertical (elevation) non-linear phase gradient is also needed to compensate for the curvature of the surface.  

\subsection{Complexity Considerations}\label{Sec:Discussion}
%

%
%
The structured phase design methods outlined in Sections \ref{subsect:structured_SE} and \ref{subsect:structured_coverage} require efficient search approaches over the space $\mathcal{S}$, whose cardinality is $|\mathcal{S}| = N^P_A$. A brute force approach is practically unfeasible even for a small number of modules, whereas the codebook size shall be dense enough to adequately match to the distribution of vehicle in the considered environment $f(\mathbf{X})$. 
Therefore, we employ a genetic algorithm (GA) for exploring $\mathcal{S}$ in \eqref{eq:SEOptimizationObjective} and \eqref{eq:CoverageOptimizationObjective}, bypassing the exhaustive grid search. In the GA context, the optimal set $\mathcal{I}$ is the most advantageous \textit{chromosome} for SE and coverage maximization, made by $P$ \textit{genes}, i.e., the number of modules. As the evaluation of the fitness function is the computational bottleneck in terms of execution time, we first pre-calculate the channel response for each angle in $\mathcal{A}_{\overline{\boldsymbol{\theta}}}$ (assuming $P=1$) and then properly combine multiple channel responses according to selected angles by $\mathcal{I}$ (for $P>1$).
\begin{figure}[t!] 
\hspace{-0.25cm}\includegraphics[width=0.55\textwidth]{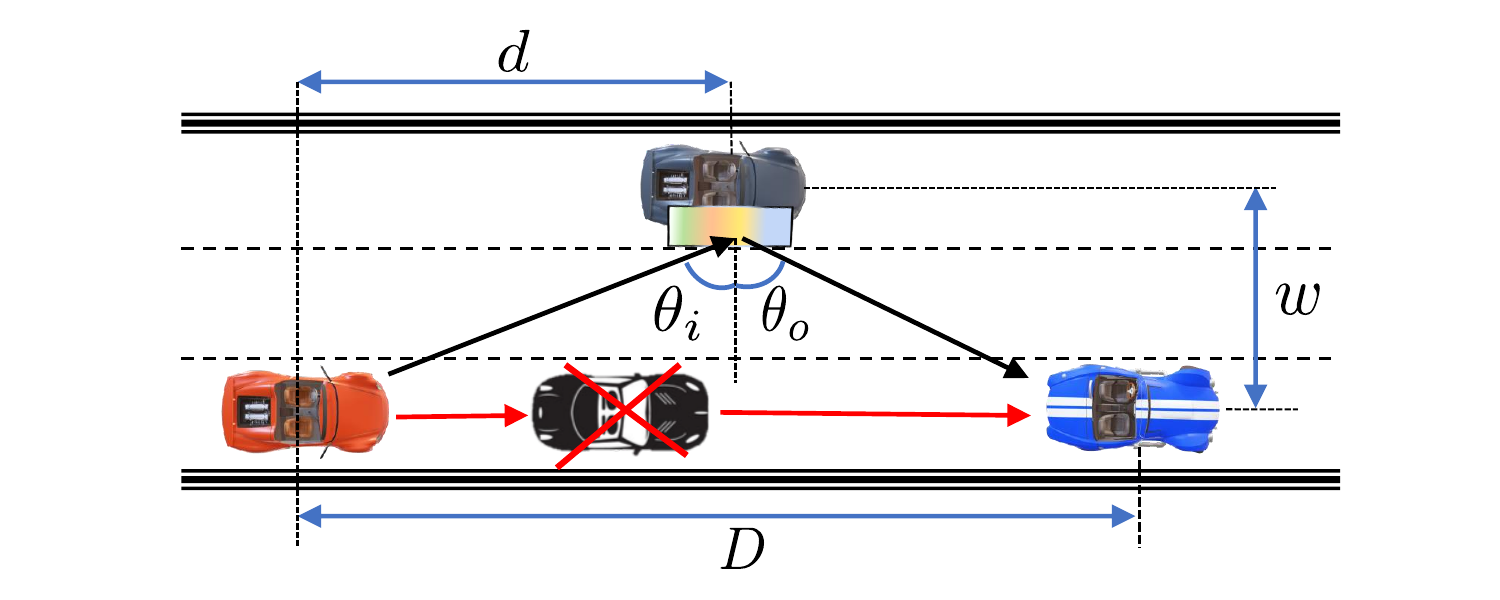}   \caption{Scenario with transmitter and receiver at the same lane, while the relaying car is located at a different lane.}\label{fig:scenario_two_lane}
\end{figure}
This operation, referred to as \textit{matrix adjustment}, boosts the GA. The codebook $\mathcal{A}_{\overline{\boldsymbol{\theta}}}$ is designed by considering AoI and AoR quantized according to the azimuth beamwidth, proportional to the number of CEMS elements along the non-curved coordinate $\Delta\theta_{bw} \propto (1/N) \simeq (1/\sqrt{L})$, where the approximation is for CEMS with $N\simeq M$. Therefore, the size of the joint AoI and AoR codebook is approximately $L$. The computational complexity of the GA applied to \eqref{eq:SEOptimizationObjective} and \eqref{eq:CoverageOptimizationObjective} can be approximated as \cite{GA_article}
\begin{equation}
\mathcal{O}\left(L^{1.5}|\mathcal{X}|G\left[\log(L) + 2\,\mu_c\,P + \mu_m\,P\right]\right)
\end{equation}
where \textit{(i)} $G$ is the number of generations, to be multiplied by the number of tested geometric configurations $|\mathcal{X}|$ (i.e., all the possible positions of Tx-Rx and CEMS), \textit{(ii)} $\mu_m$ and  and $\mu_c$ are the mutation and cross-over percentages, respectively \cite{GA_article}. $\mu_m$ is usually very low, thus the last term can be neglected. The number of modules $P$ is linearly affecting the computation time, while the complexity of a brute force approach scales exponentially with $P$. Note the reduction in complexity w.r.t. unstructured approaches \eqref{eq:PenaltyModelSNR} and \eqref{eq:PenaltyModelSNR3} (Section \ref{subsect:unstructured_complexity}).

\section{Numerical Results}\label{sect:numerical_results}

\subsection{Structured vs. Unstructured Phase Design}\label{Sec:Sim1}

\begin{table}[!b]
    \centering
    \footnotesize
    \caption{Simulation parameters used in Section \ref{Sec:Sim1}.}
    \begin{tabular}{l|c|c}
    \toprule
        \textbf{Parameter} &  \textbf{Symbol} & \textbf{Value(s)}\\
        \hline
        Carrier frequency & $f_0$  & $28$ GHz \\
        Bandwidth & $B$ & $200$ MHz\\
        Transmitted power & $\sigma^2_s$ & $23$ dBm\\
        Noise power & $\sigma^2_n$ & $-82$ dBm\\
        CEMS elements & $N \times M$ & $24 \times 12$\\
        CEMS elements's spacing & $d_n,d_m$ & $\lambda_0/4$ m\\
        CEMS radius of curvature & $R$ & 2 m\\
        Number of CEMS modules & $P$ & varying \\
        Tx and Rx elements & $N_\mathrm{Tx}, N_\mathrm{Rx}$ & $8,8$ \\
        Tx and Rx elements's spacing & $d_\mathrm{Tx}, d_\mathrm{Rx}$ & $\lambda_0/2$ m \\
        Codebook size & $N_A$ & 64\\
        Car density (per lane) & $\lambda_{car}$ & 30 cars/Km\\
        \bottomrule
    \end{tabular}
    \label{tab:SimParam_1}
\end{table}

\begin{figure*}[t!]
    \centering
    \subfloat[]{\includegraphics[width=0.45\textwidth]{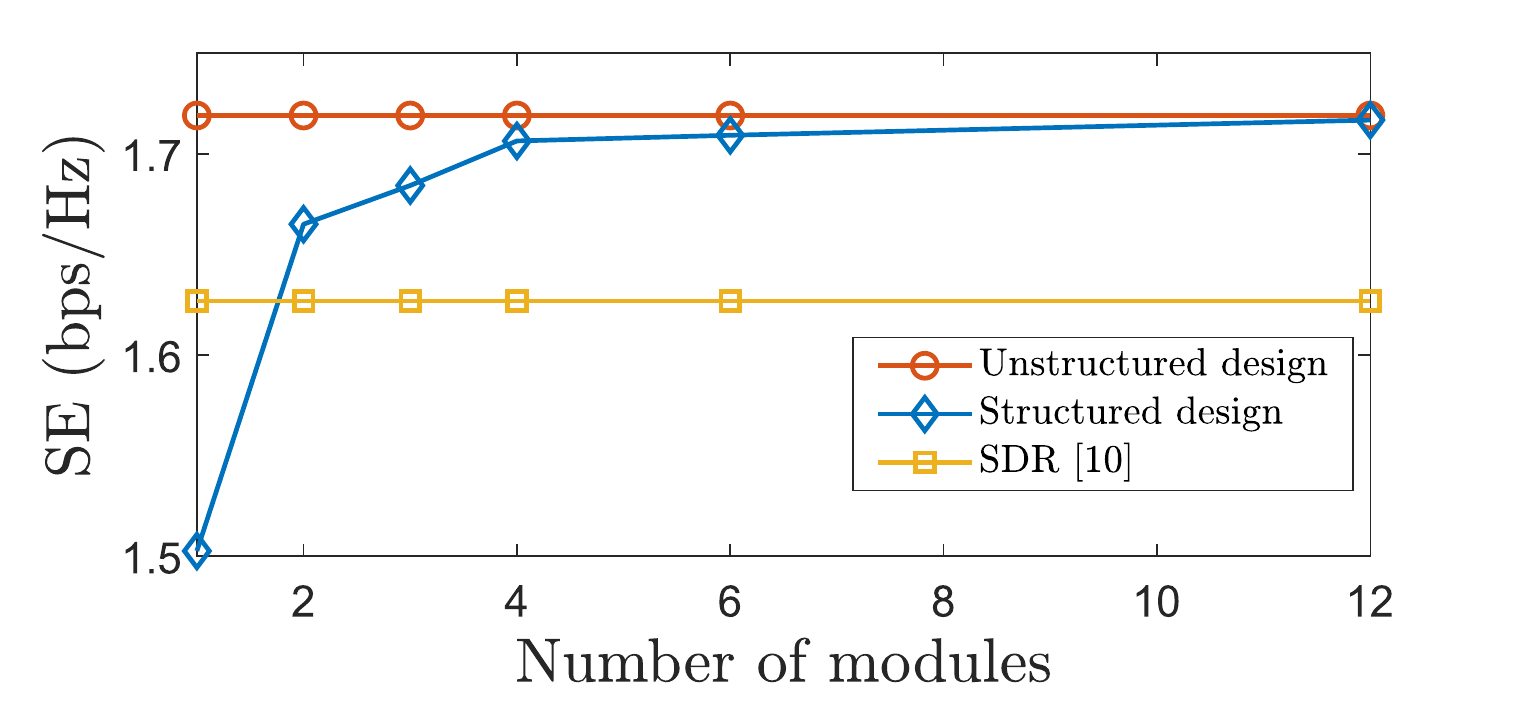}}\quad
    \subfloat[]{\includegraphics[width=0.45\textwidth]{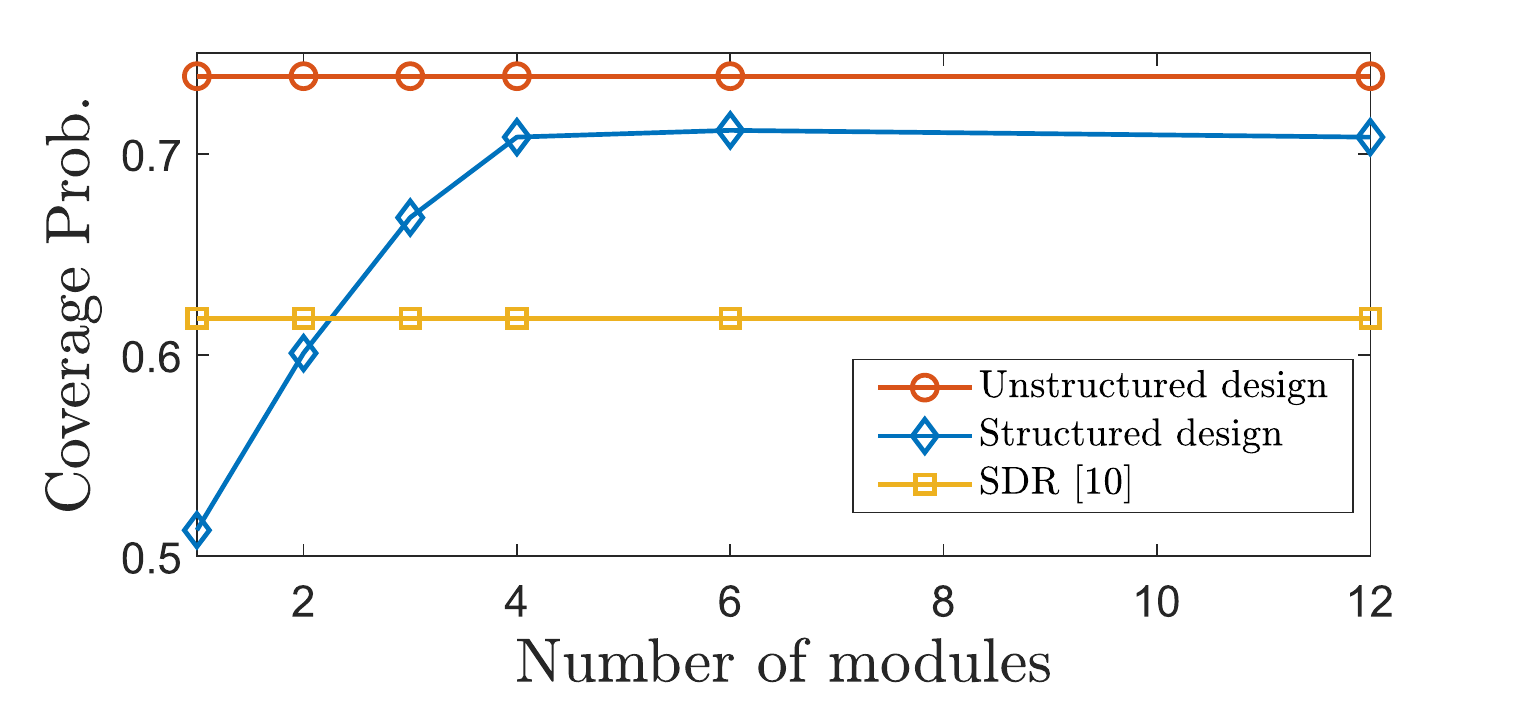}}
    \caption{Optimization results in terms of (a) average SE, and (b) average coverage probability. The coverage threshold is set to be  $\gamma = 0$ dB.}\label{fig:Sim1}
\end{figure*}
\begin{figure*}[t!]
    \centering
    \subfloat[]{\includegraphics[width=0.32\textwidth]{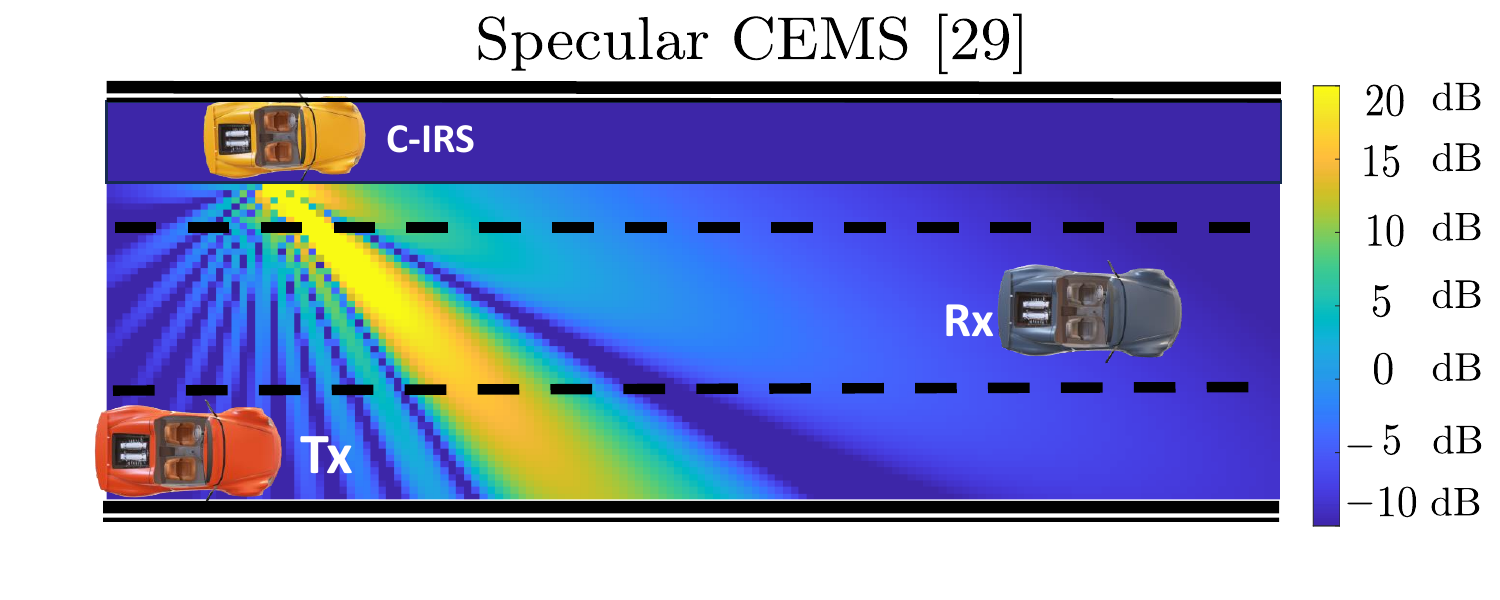}}\quad
    \subfloat[]{\includegraphics[width=0.32\textwidth]{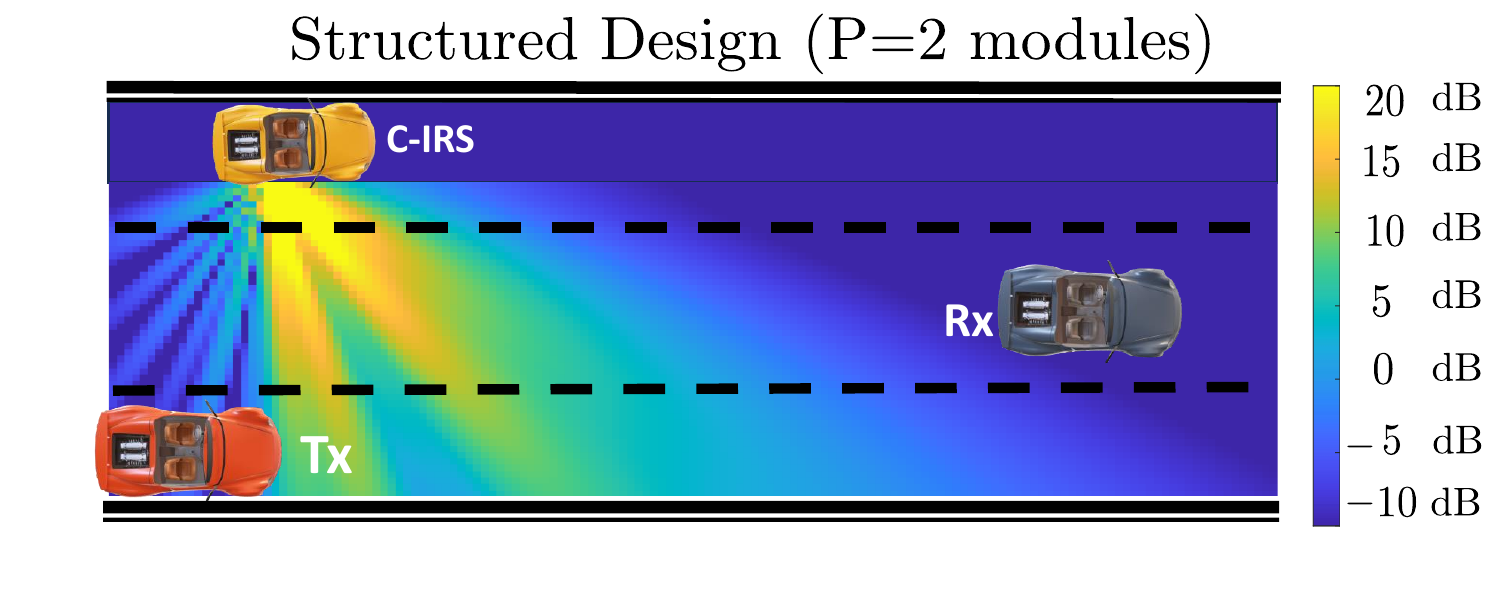}}\quad
    \subfloat[]{\includegraphics[width=0.32\textwidth]{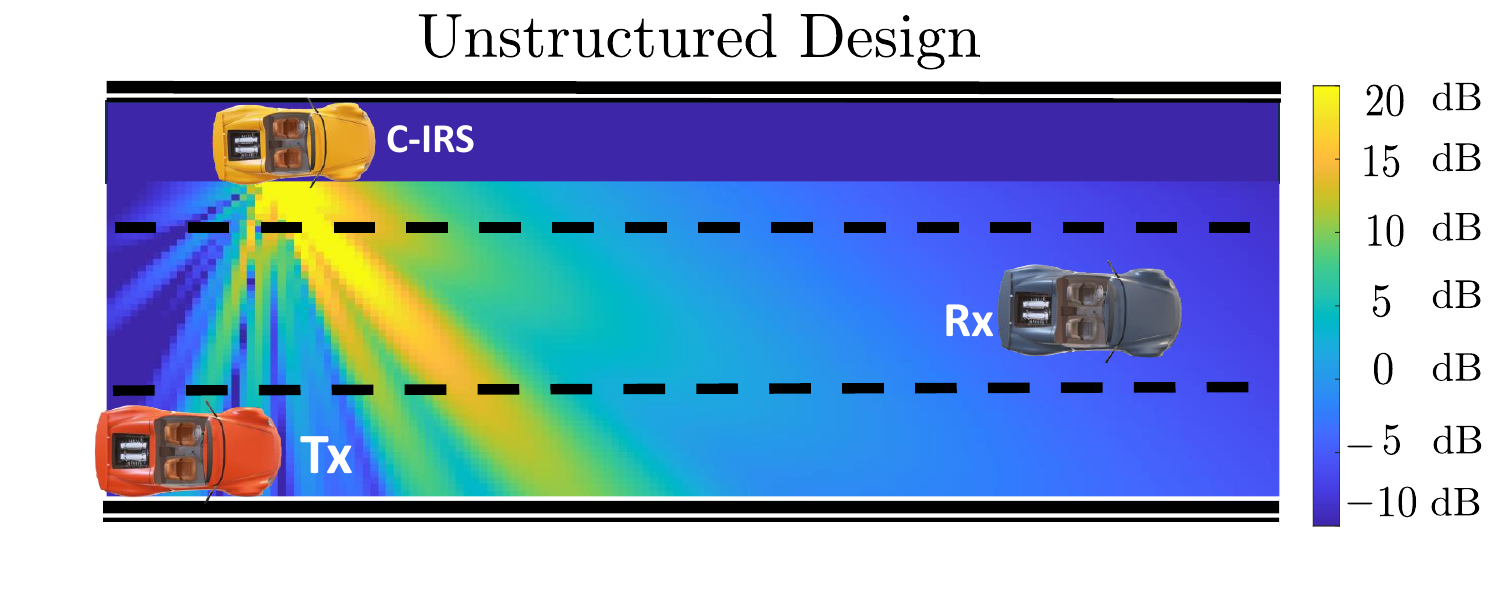}}
    \caption{Map of the SNR in dB for (a) specular reflection \cite{CIRS_TWC}, (b) anomalous reflection, employing the structured optimization method in Section \ref{sec:modular_design} ($P=2$ modules), and (c) anomalous reflection, employing the unstructured penalty optimization method in Section \ref{sec:Opt_Unstructured}. The Tx and relay positions are $\mathbf{x}_\mathrm{Tx} = [20,0,2]^T$ and $\mathbf{x}_\mathrm{C}= [0,20,2]^T$, while the Rx is let vary to build the SNR map.}\label{fig:HeatMap}
\end{figure*}


The first set of results compare the proposed structured and unstructured optimization methods for the design of the CEMS, for a two lane scenario. Tx and Rx are located on the same lane (direct lane) at the same height of the relaying vehicle equipped with CEMS, the latter situated on the second lane (relay lane), as depicted in Fig. \ref{fig:scenario_two_lane}. The statistical distribution of AoI and AoR $\theta_i$ and $\theta_o$ is obtained from the random distribution of Tx, Rx, and relay over their pertaining lane. Tx and Rx are distributed according to a Poisson line process~\cite{V2V_Blockage}, where their mutual distance $D$ follows a negative exponential distribution with parameter $\lambda_{car} = 30$ cars/Km (vehicles' spatial density). The minimum generated Tx-Rx distance is $30$ m, as for shorter distances assuming blockage may be not realistic. The distance between Tx, Rx, and the relay $d$ is instead uniformly distributed from $0$ to $D$. The AoI and AoR are therefore $\theta_i = \arctan (d/w)$ and $\theta_o = \arctan((D-d)/w)$, where $w$ is the inter-lane distance. The list of simulation parameters used for this comparison is summarized in Table \ref{tab:SimParam_1}. As unstructured optimization methods are impractical for configuring large-sized surfaces due to computational complexity, we compare the two methods for a $N \times M = 24 \times 12$ CEMS.   

Fig.~\ref{fig:Sim1} shows the results of optimization in terms of average SE (Fig.~\ref{fig:Sim1}a), and average coverage probability (Fig.~\ref{fig:Sim1}b). The proposed penalty for unstructured optimizations outperforms the classical \gls{SDR} method \cite{OPT_SNR_1}. This is due to the fact that SDR searches for a rank 1 solution of matrix $\mathbf{V}$, that is usually not guaranteed and a further randomization step is necessary, yielding sub-optimal results~\cite{Zhang_Randomization}. On the other hand, the structured optimization with a modular architecture with P modules and much less computation time, can achieve similar performance to the unstructured penalty method, if the number of modules $P$ is large enough (here $p \geq6$). The configurations of the \gls{CEMS} obtained in this stage are utilized to generate exemplary maps of the SNR, illustrated in Fig. \ref{fig:HeatMap}, for: \textit{(i)} specular reflection \cite{CIRS_TWC} (Fig. \ref{fig:HeatMap}a), \textit{(ii)} anomalous reflection, employing the structured optimization method in Section \ref{sec:modular_design} (2 modules, Fig. \ref{fig:HeatMap}b) \textit{(iii)} anomalous reflection, employing the unstructured penalty optimization method in Section \ref{sec:Opt_Unstructured} (Fig. \ref{fig:HeatMap}c). The Tx and relay positions are $\mathbf{x}_\mathrm{Tx} = [20,0,2]^T$ and $\mathbf{x}_\mathrm{C}= [0,20,2]^T$, while the Rx is let vary to build the SNR map. As expected, a single specular reflection (from \cite{CIRS_TWC}) allows for effectively covering a small set of AoI/AoR, while both the proposed optimization methods lead to remarkable improvements.

\subsection{Structured Phase Design for Large CEMS} \label{Sec:Sim2}

\begin{table}[!b]
    \centering
    \footnotesize
    \caption{Simulation parameters used in Section \ref{Sec:Sim2}.}
    \begin{tabular}{l|c|c}
    \toprule
        \textbf{Parameter} &  \textbf{Symbol} & \textbf{Value(s)}\\
        \hline
        Number of lanes & -  & 2 \\
        Carrier frequency & $f_0$  & $28$ GHz \\
        Bandwidth & $B$ & $200$ MHz\\
        Transmitted power & $\sigma^2_s$ & $23$ dBm\\
        Noise power & $\sigma^2_n$ & $-82$ dBm\\
        CEMS elements & $N \times M$ & varying\\
        CEMS elements's spacing & $d_n,d_m$ & $\lambda_0/4$ m\\
        CEMS radius of curvature & $R$ & 2 m\\
        Number of CEMS modules & $P$ & varying \\
        Tx and Rx elements & $N_\mathrm{Tx}, N_\mathrm{Rx}$ & $8,8$ \\
        Tx and Rx elements's spacing & $d_\mathrm{Tx}, d_\mathrm{Rx}$ & $\lambda_0/2$ m \\
        Codebook size & $N_A$ & varying\\
        Car density (per lane) & $\lambda_{car}$ & 30 cars/Km\\
        \bottomrule
    \end{tabular}
    \label{tab:SimParam_2}
\end{table}

The second set of results concerns the SE and coverage analysis in the same two-lane scenario of Fig. \ref{fig:scenario_two_lane}, considering more realistic large-sized \gls{CEMS}. 
\begin{figure}[!t]
    \centering
    \subfloat[][]{\includegraphics[width=0.45\textwidth]{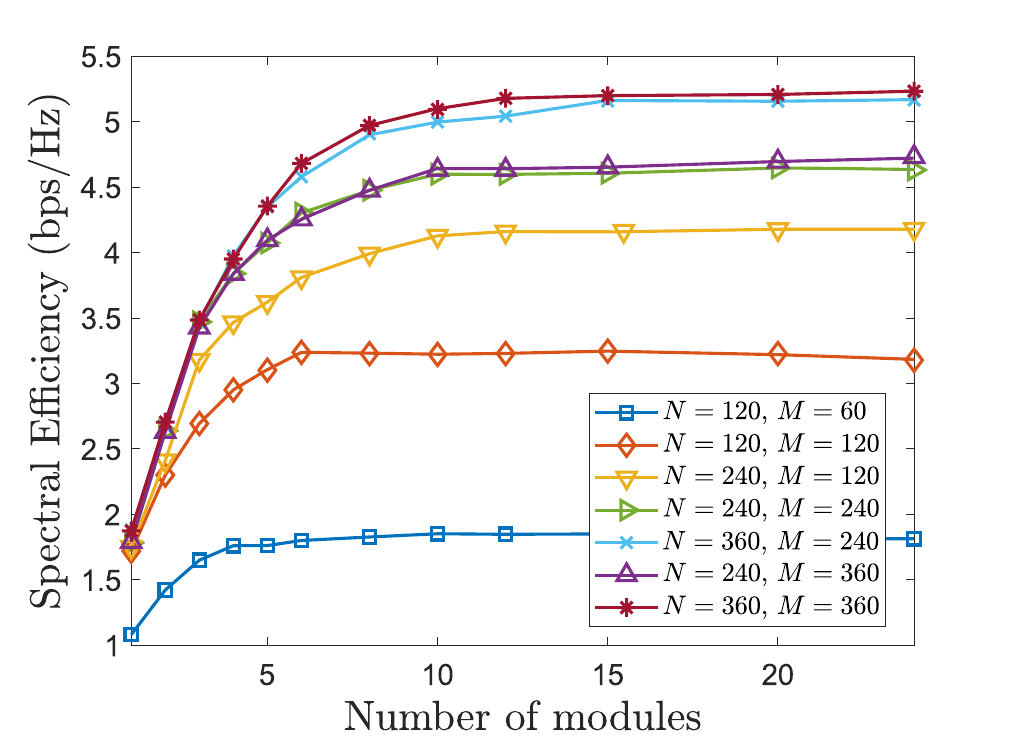}\label{fig:SE_Comparison}}\\
    \subfloat[][]{\includegraphics[width=0.45\textwidth]{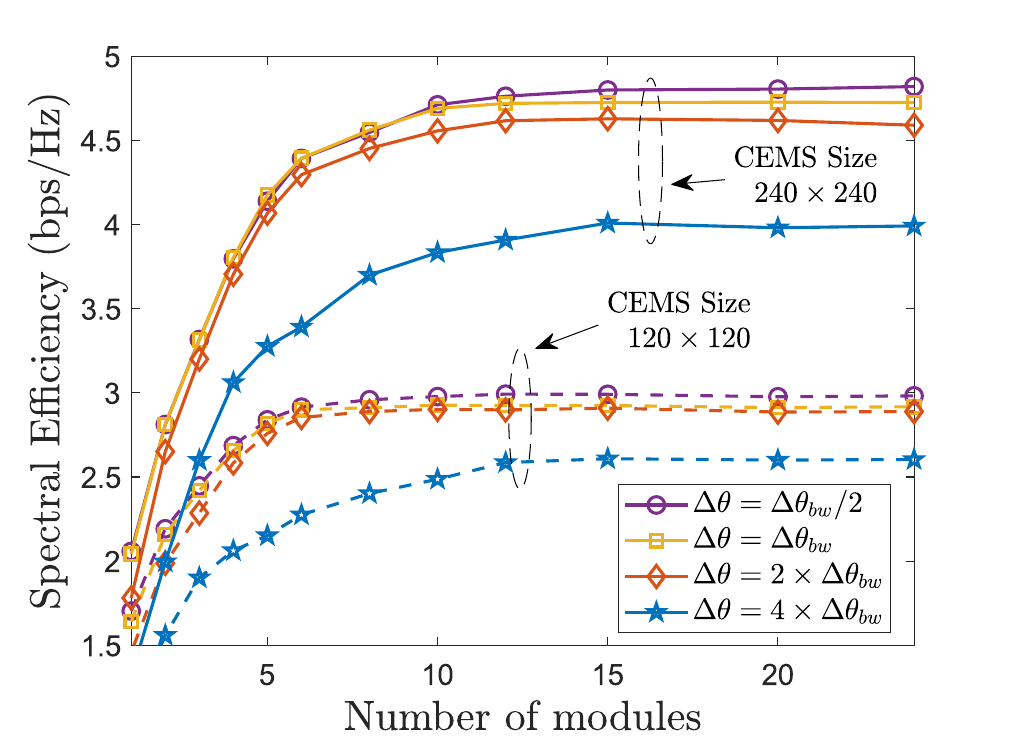}\label{fig:CodeBookSize}}\\
    \caption{SE vs. number of modules $P$ for CEMS designed with the structured method (Section \ref{sec:modular_design}): (a) varying CEMS size and (b) varying the codebook size. Two CEMS dimensions with $240\times 240$ elements, and $120\times120$ elements are shown.}
    \label{fig:SE_results_large}
\end{figure}
\begin{figure}[t!] 
    \centering
\includegraphics[width=0.45\textwidth]{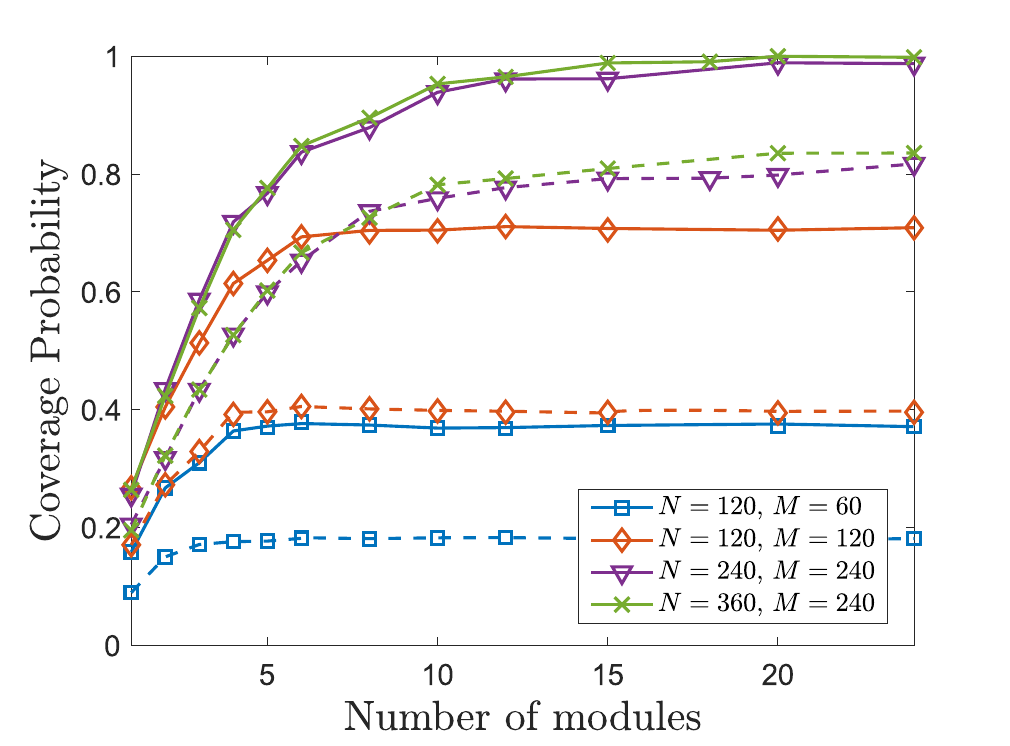}
    \caption{ Average coverage probability vs. number of modules $P$ for CEMS designed with the structured method (Section \ref{sec:modular_design}), varying the CEMS size. The coverage SNR threshold is set to $\gamma = 5$ dB for solid curves, and  $\gamma =10$ dB for dashed curves.}
    \label{fig:Cov_Comparison}
\end{figure}
Fig. \ref{fig:SE_Comparison} illustrates the average \gls{SE} versus the number of CEMS modules $P$, varying \gls{CEMS} size. Simulation parameters are summarized in Table \ref{tab:SimParam_2}. Depending on the \gls{CEMS} size, between $P=5$ to $P=10$ modules are necessary to guarantee the adequate spatial diversity in the anomalous reflection,  yielding a SE that saturates at its maximum value, that is approximately three times the achieved SE for $P=1$ \cite{CIRS_TWC}. Notably, increasing the size of the \gls{CEMS} from $120\times 60$ to $120 \times 120$ results in a substantial gain in \gls{SE}, while passing from $360 \times 240$ to $360\times 360$ gives marginal benefit. Indeed, the SE performance (as well as the coverage one) is mostly ruled by the azimuth resolution, i.e., by the number of elements $N$ along the cylindrical coordinate. This latter insight is confirmed by the fact that a \gls{CEMS} with dimensions $240 \times 240$ performs equivalently to $240 \times 360$, while a \gls{CEMS} with dimensions of $360\times 240$ outperforms both. Thus, increasing the CEMS size along the curved coordinate (number of elements $M$) is not much beneficial for very large-sized CEMS. In general, choosing the aspect ration of the CEMS is a crucial task that depends on the specific performance metric. 

The impact of the angular codebook size $N_A$ is analyzed in Fig. \ref{fig:CodeBookSize}, showing the average SE versus the number of modules $P$, varying $N_A$. \gls{CEMS} size is set to $120\times 120$ (dashed lines) and $240\times 240$ (solid lines). The sampling interval over the azimuth domain is set as function of the null-to-null beam width $\Delta\theta_{\text{bw}} \simeq 1/N$. Fig. \ref{fig:CodeBookSize}  highlights that exceeding $\Delta\theta_{\text{bw}}$, i.e., $\Delta \theta > \Delta \theta_{bw}$ generally decrease the performance, while increasing the granularity, i.e., $\Delta \theta < \Delta \theta_{bw}$ improves the performance. In this specific case, choosing $\Delta \theta = 4 \Delta \theta_{bw}$ makes the SE decrease. The extent of the latter deterioration increases with the CEMS size.

As last, Fig. \ref{fig:Cov_Comparison} presents the average coverage probability under the same configurations of Fig. \ref{fig:SE_Comparison}. We examine two cases of coverage SNR thresholds, $\gamma = 5$ dB and $\gamma = 10$ dB. Similarly to the insight obtained for the SE, between $P=5$ to $P=15$ modules are necessary to the saturation of the coverage probability. With a sufficient number of modules, \gls{CEMS}s can significantly enhance the average coverage probability by approximately 20\% (comparably small CEMS) to 100\% (comparably large CEMS).

\subsection{Realistic Multi-Lane Scenario with Structured Optimization} \label{Sec:Sim3}

\begin{table}[!b]
    \centering
    \footnotesize
    \caption{Simulation parameters used in Section \ref{Sec:Sim3}.}
    \begin{tabular}{l|c|c}
    \toprule
        \textbf{Parameter} &  \textbf{Symbol} & \textbf{Value(s)}\\
        \hline
        Number of lanes & -  & 4 \\
        Carrier frequency & $f_0$  & $28$ GHz \\
        Bandwidth & $B$ & $200$ MHz\\
        Transmitted power & $\sigma^2_s$ & $23$ dBm\\
        Noise power & $\sigma^2_n$ & $-82$ dBm\\
        CEMS elements & $N \times M$ & $360\times 240$\\
        CEMS elements's spacing & $d_n,d_m$ & $\lambda_0/4$ m\\
        CEMS radius of curvature & $R$ & 2 m\\
        Number of CEMS modules & $P$ & 24 \\
        Tx and Rx elements & $N_\mathrm{Tx}, N_\mathrm{Rx}$ & $8,8$ \\
        Tx and Rx elements's spacing & $d_\mathrm{Tx}, d_\mathrm{Rx}$ & $\lambda_0/2$ m \\
        Codebook size & $N_A$ & 2209\\
        Car density (per lane) & $\lambda_{car}$ & $10,60$ cars /km\\
        \bottomrule
    \end{tabular}
    \label{tab:SimParam_3}
\end{table}

The last set of results shows the benefits of employing CEMS in a more realistic 4-lane vehicular scenario. On each lane, vehicles are distributed according to a Poisson line process, as explained in Section \ref{Sec:Sim1}. Each vehicle is equipped with CEMS on both car doors, CEMS being composed of $360 \times 240$ elements, divided in $P=24$ modules. Tx and Rx vehicles are randomly chosen among the generated vehicles, and the pdf of the azimuth AoI and AoR $\theta_i$ and $\theta_o$ is obtained by Monte Carlo sampling over the vehicles' Poisson distribution, namely $\mathbf{x}_\mathrm{Tx}, \mathbf{x}_\mathrm{Rx}$ and all the candidate relaying vehicles $\{\mathbf{x}_\mathrm{C}\}$, whose number and locations change at each generation. The CEMS are optimized for SE using the structured method of Section \ref{subsect:structured_SE}. Note that to generate the following figures, the CEMS are optimized for SE. The same analysis can be done with CEMS optimized for coverage. However, due to similarity of the final results, the latter is not shown for brevity.\\ 
After the CEMS design, another Monte Carlo simulation is carried out where the blockage condition between the Tx and Rx, Tx and relay, relay and Rx is tested for each of the candidate relays in the considered area. If at least one between the Tx-relay and relay-Rx is blocked, that specific relay is removed from the candidate list. Then, if the Tx-Rx direct path is also blocked, the communication takes places following either \textit{(i)} random relay selection among the remaining candidate ones or \textit{(ii)} the Tx selects the relay that maximizes the Rx power. It's important to note that in both cases, the Tx and Rx are assumed to be aware of the relay's position, allowing for beamforming and precoding adjustments accordingly. Simulation parameters are summarized in Table \ref{tab:SimParam_3}.
\begin{figure}
    \centering
    \subfloat[][]{\includegraphics[width=0.45\textwidth]{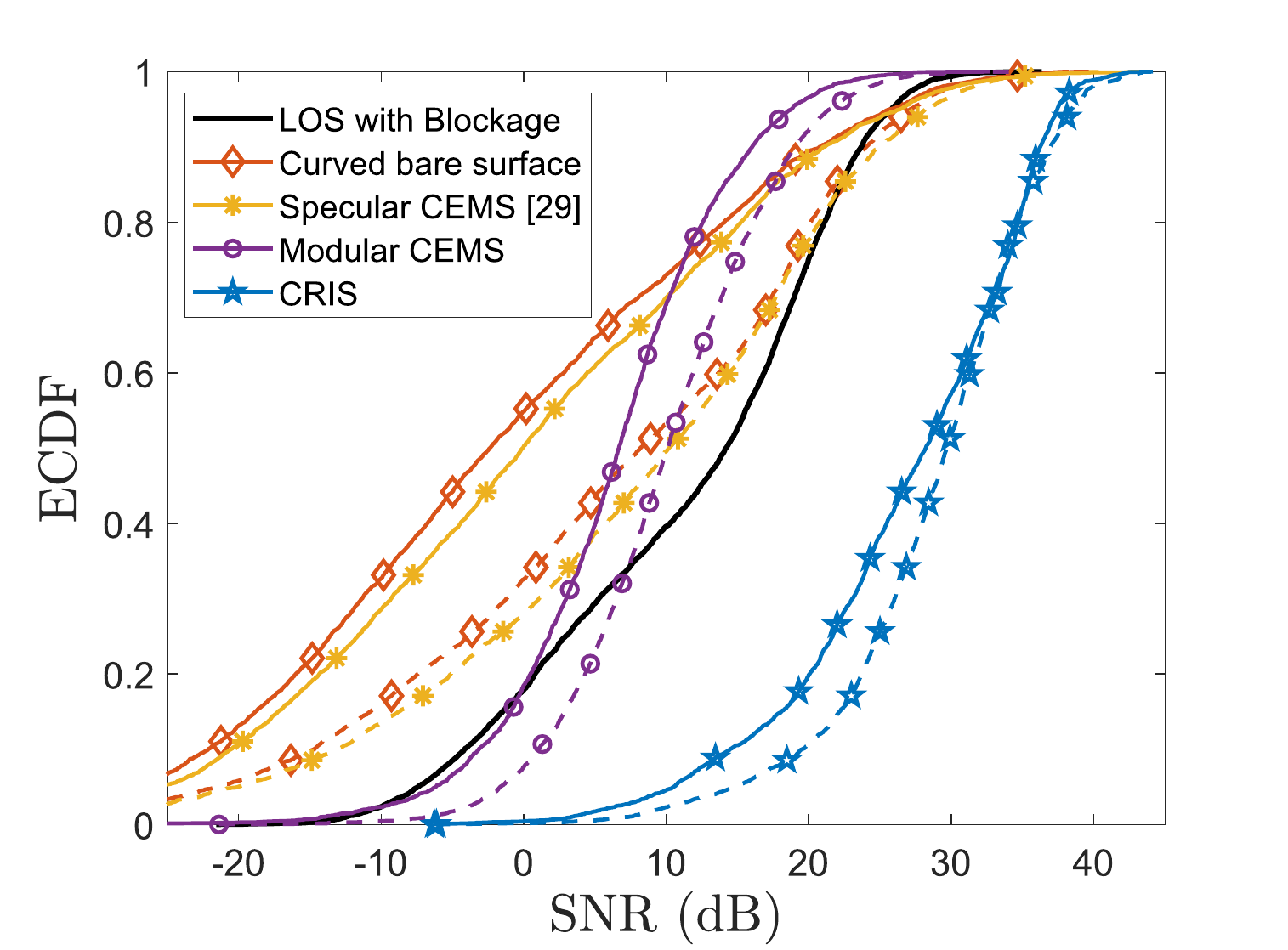}\label{fig:ECDF1}}\\
    \subfloat[][]{\includegraphics[width=0.45\textwidth]{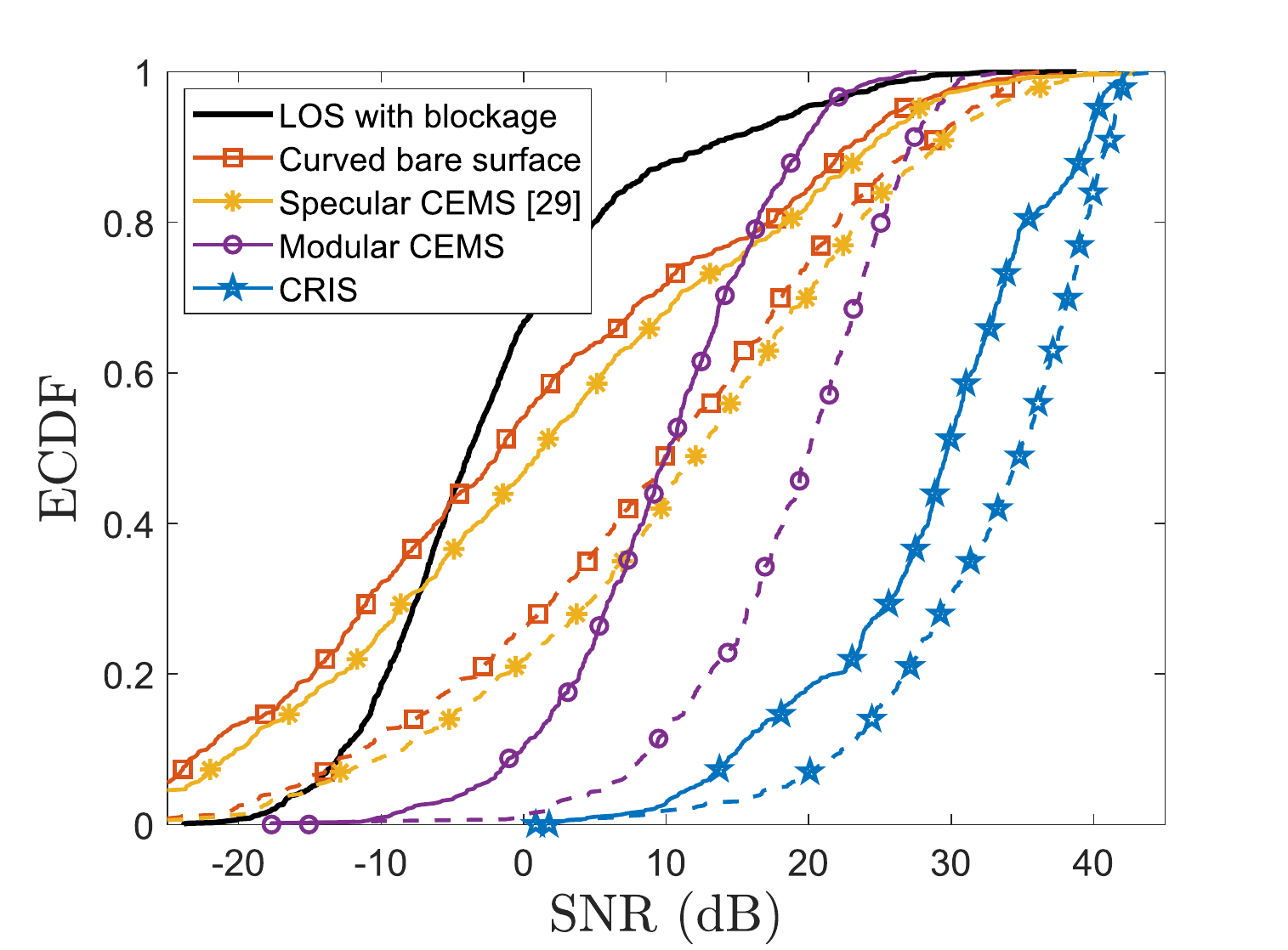}\label{fig:ECDF2}}
    \caption{ECDF of SNR obtained by using either the direct link or the relayed link, where the relaying car is equipped with CEMS or CRIS and (a) low traffic condition, $\lambda_{car} = 10$ cars/Km, (b) high traffic condition $\lambda_{car} = 60$ cars/km. Solid lines correspond to random relay selection and dashed lines correspond to maximum selection among possible relays.}
    \label{fig:ECDF}
\end{figure}
Figs. \ref{fig:ECDF1} and \ref{fig:ECDF2} show the empirical cumulative density function (ECDF) of the SNR, for a CEMS with size $360 \times 240$, with the density of vehicles set as $\lambda_{car} = 10$ cars/km (Fig. \ref{fig:ECDF1}) and $\lambda_{car} = 60$ cars/km (Fig. \ref{fig:ECDF2}). We consider 5 different cases: \textit{(i)} the Tx uses only the possibly blocked direct link with RX (solid black line), \textit{(ii)} the Tx uses the bare car door of the relay for specular reflection (red lines) \textit{(iii)} the relay is equipped with a CEMS designed to "flatten" the car door acting for specular reflection~\cite{CIRS_TWC}, \textit{(iv)} the relay is equipped with a modular CEMS designed with the proposed structured method in Section \ref{subsect:structured_SE}, \textit{(v)} the relay is equipped with an optimized curved RIS (CRIS). Solid lines refer to the random selection of the relay, while dashed lines indicate the maximum power selection. The lines for curved RIS are reported as an upper-bound benchmark, obtained by setting $\phi_{\ell}(\overline{\boldsymbol{\vartheta}}) = \phi_{\ell}(\boldsymbol{\vartheta})$ in \eqref{eq:phase_config} (using the true AoI and AoR)\footnote{In practice, using such a reconfigurable surface with a huge number of elements such as $360\times 240$ is impractical, due to high cost and periodic channel estimation overhead imposed by channel mobility \cite{haghshenas2023parametric}}. The specular CEMS is instead designed to compensate the curvature of the car door for the azimuth $\theta_o = - \theta_i = 76$ deg~\cite{CIRS_TWC}. In a low traffic condition, Fig. \ref{fig:ECDF1}, the blockage probability of the direct path is low, thus the usage of the direct link provides comparable performance to the usage of a any relayed link with CEMS. As expected, the worst possible performance is represented by the usage of the bare surface of the car door, as most the energy is not reflected in the Rx direction. A similar conclusion can be drawn for specular CEMS, as the candidate relays can be found in a very limited section (see Fig. \ref{subfig:scenario_specular}). Exmploying optimized modular CEMS improves the performance, but, on average, there is no significant advantage w.r.t. using the direct link. In contrast, CRIS provide the upper performance bound as, in principle, \textit{any} vehicle can be used as a relay.

In high traffic conditions (Fig. \ref{fig:ECDF2}), the advantages of using CEMS becomes relevant. Indeed, the blockage probability of the direct path strongly increases, limiting its reliability. Therefore, the direct path exhibits the poorest performance, overall. In contrast, the possibility of using one of the CEMS-equipped relays provides a valid alternative in coverage augmentation, outperforming the usage of the direct path only. Remarkably, in coverage there is a clear advantage in employing the proposed optimized modular CEMS w.r.t. a single specular surface \cite{CIRS_TWC}, as it can be inferred by inspection of the tail of the ECDFs. While a single specular CEMS can attain high \textit{peak values} of SNR (that occur for relays perfectly placed halfway between Tx and Rx), a modular approach increases the \textit{weighted average} of the SNR. 
Therefore, the usage of static passive CEMS not only significantly reduces costs w.r.t. CRIS, but it also eliminates the complexities associated with channel estimation and dynamic adjustments, thus making large-scale implementations feasible.
\section{Conclusion} \label{sect:conclusion}

 Our design criteria of curved EMSs for vehicular networks reveal a remarkable enhancements in communication performance. By leveraging low-cost, static passive CEMSs mounted on car doors, we demonstrate through theoretical analysis and numerical simulations that smart pre-configuration can substantially mitigate signal blockage issues. The optimization of the CEMS phase profile, via both unstructured and structured design methods, results in considerable improvements in spectral efficiency and coverage probability. Specifically, our results show that the structured design of a modular CEMS architecture can achieve near-benchmark performance with reduced computational complexity, offering a viable solution for enhancing vehicular network communication in challenging environments. These advancements underscore the potential of CEMSs to serve as effective tools for improving the reliability and efficiency of future vehicular communications.

\bibliographystyle{IEEEtran}
\bibliography{main}

\end{document}